%% file: cse3.tex
\documentclass[acmsmall]{acmart}

\newif\ifarxiv
\arxivtrue   

\newcommand{\appref}[2]{%
  \ifarxiv
    \Cref{#1}\apprefcheck{#1}{#2}%
  \else
    the extended paper~\cite[App.~#2]{cse3-extended}%
  \fi}

\newcommand{\apprefcheck}[2]{%
  \ifcsname r@#1\endcsname
    \edef\apprefactual{\getrefnumber{#1}}%
    \edef\apprefgiven{#2}%
    \ifx\apprefactual\apprefgiven\else
      \ClassError{cse3}
        {Appendix `#1' is numbered \apprefactual, but \string\appref\space says #2}
        {Change the second argument of that \string\appref\space to \apprefactual.}%
      \batchmode \read -1 to \apprefactual  
    \fi
  \fi}

\hypersetup{hypertexnames=false}

\setcopyright{cc}
\setcctype{by}
\acmDOI{10.1145/3871744}
\acmYear{2027}
\acmJournal{PACMPL}
\acmVolume{11}
\acmNumber{POPL}
\acmArticle{42}
\acmMonth{1}
\acmSubmissionID{popl27main-p345-p}
\ifarxiv\else 
\received{2026-07-09}
\received[accepted]{2026-10-05}
\fi

\usepackage{thmtools}

\usepackage[utf8]{inputenc}
\usepackage{amsmath}
\usepackage{tabularx}
\usepackage{booktabs}
\usepackage{multirow}

\usepackage{tikz}
\usetikzlibrary{arrows, arrows.meta}
\usetikzlibrary{calc}

\usepackage{wrapfig}
\usepackage{subcaption}

\usepackage{tabularx}

\usepackage{graphicx}
\graphicspath{ {./images/} }

\usepackage{xspace}
\usepackage{annotate-equations}
\usepackage[nameinlink]{cleveref}

\usepackage{listings}
\usepackage{lstocaml}
\usepackage{listings-rocq}

\usepackage{CSE3macros}
\usepackage{sl}
\usepackage{c_macros}
\usepackage[inline]{enumitem}
\setlist[itemize]{leftmargin=21pt}

\renewcommand{\implies}{\Rightarrow}
\renewcommand{\iff}{\Leftrightarrow}

\AtBeginEnvironment{example}{\setlength{\parindent}{0pt}}
\AtBeginEnvironment{definition}{\setlength{\parindent}{0pt}}

\begin{document}

\ifarxiv
  \title[Compositional Symbolic Execution à la Carte (Extended Version)]
        {Compositional Symbolic Execution à la Carte\texorpdfstring{\\}{ }(Extended Version)}
\else
  \title{Compositional Symbolic Execution à la Carte}
\fi

\author{Sacha-Élie Ayoun}
\correspondingauthor
\authornote{Both authors contributed equally to this research}
\orcid{0000-0001-9419-5387}
\affiliation{%
  \institution{Imperial College London}
  \city{London}
  \country{United Kingdom}}
\email{s.ayoun17@imperial.ac.uk}

\author{Andreas Lööw}
\authornotemark[2] 
\orcid{0000-0002-9564-4663}
\affiliation{%
  \institution{Royal Holloway, University of London}
  \city{Egham}
  \country{United Kingdom}}
\email{andreas.loow@rhul.ac.uk}

\author{Seung Hoon Park}
\orcid{0000-0001-7165-6857}
\affiliation{%
  \institution{Imperial College London}
  \city{London}
  \country{United Kingdom}}
\email{s.park23@imperial.ac.uk}

\author{Opale Sjöstedt}
\orcid{0009-0003-7545-1383}
\affiliation{%
  \institution{Imperial College London}
  \city{London}
  \country{United Kingdom}}
\email{opale.sjostedt23@imperial.ac.uk}

\author{Petar Maksimović}
\orcid{0000-0002-0400-7467}
\affiliation{%
  \institution{Harmonic}
  \city{London}
  \country{United Kingdom}}
\email{petar.maksimovic@harmonic.fun}

\author{Philippa Gardner}
\orcid{0000-0002-4187-0585}
\affiliation{%
  \institution{Imperial College London}
  \city{London}
  \country{United Kingdom}}
\email{p.gardner@imperial.ac.uk}



\begin{abstract}
We introduce compositional symbolic execution à la carte, a modular approach for building compositional symbolic execution (CSE) tools, powered by combinator-based design of key components, including memory models, management and automation of user-defined predicates, and bi-abductive analysis. Our approach is grounded on meta-theory mechanised in the Rocq proof assistant and allows tool developers to easily build correct-by-construction CSE tools for symbolic analysis of their programming languages of choice, with minimal implementation and proof effort.
We show how our approach scales to real-world languages by implementing it in Gillian, a state-of-the-art CSE tool, and using it to capture the intricacies of the memory models of C, JavaScript, and Rust.
We evaluate our approach by running all of the relevant previous case studies of Gillian, showing no unexpected behaviours and comparable performance.
\end{abstract}

\maketitle

\input{sections/introduction}
\input{sections/overview}
\input{sections/background}
\input{sections/zoo}
\input{sections/eval}
\input{sections/limitations}
\input{sections/related-work}
\input{sections/conclusion}

\bibliography{cse3.bib}

\ifarxiv
\appendix
\crefalias{section}{appendix}
\crefalias{subsection}{appendix}
\include{sections/app-zoo}

\include{sections/app-eval}
\else
\fi

\end{document}

%% file: sections/introduction.tex

\section{Introduction}
\label{sec:intro}

\emph{Compositional symbolic execution} (CSE)~\cite{compositional-symbolic-execution-loow-2024} is a broadly applied program analysis technique used in state-of-the-art verification and bug-finding tools such as VeriFast~\cite{verifast-powerful-sound-jacobs-2011}, Viper~\cite{viper-verification-infrastructure-muller-2016}, Gillian~\cite{gillian-part-multilanguage-fragososantos-2020,gillian-part-ii-maksimovic-2021,compositional-symbolic-execution-loow-2024}, CN~\cite{cn-verifying-systems-pulte-2023}, and Infer-Pulse~\cite{finding-real-bugs-le-2022,unter}.
CSE tools perform symbolic execution grounded on separation logic (SL)~\cite{seplogic,separation-logic-logic-reynolds-2002} and its variants, such as incorrectness SL (ISL)~\cite{local-reasoning-presence-raad-2020} and implicit dynamic frames~\cite{implicit-dynamic-frames-smans-2012}.
This use of SL enables \emph{compositional} reasoning, where functions are analysed in isolation and specifications are reused across different contexts, allowing CSE tools to handle heap-manipulating programs and scale to large codebases.

The implementation and extension of CSE engines for real-world programming languages are costly and highly complex,
as one must both accurately represent the state of the analysed language, including its memory model (i.e.,~heap), and incorporate ghost state needed for the analysis being performed. This ghost state can range from, e.g., simple fractional permissions for shared data~\cite{fractional-permissions-boyland-2013} or markers for tracking deallocated memory cells~\cite{local-reasoning-presence-raad-2020}, to bespoke, application-specific structures such as those required for analysing real-world Rust~\cite{rustbelt-securing-foundations-jung-2017,rusthornbelt-semantic-foundation-matsushita-2022,hybrid-approach-semiautomated-ayoun-2025}.
The design choices a tool developer makes when designing such symbolic states (e.g., when and how the engine should branch when reading and updating state) are shaped by various trade-offs involving scalability, expressivity, and performance. Moreover, formal soundness proofs of CSE engines are notoriously difficult to maintain, especially if design decisions are changing underfoot.

Our primary goal is to \emph{relieve tool builders of the implementation and proof burden associated with designing CSE memory models}. 
To this end, we present the first \emph{modular approach to CSE}, which we call \emph{CSE à la carte}, making two main contributions:
\begin{enumerate}
\item we introduce a set of reusable components that capture a broad range of program analysis concepts and that facilitate the design of correct-by-construction CSE tools for both correctness and incorrectness reasoning about real-world programming languages;
\item we provide two complementary implementations of our approach, a foundational one for the meta-theory in the Rocq proof assistant and an execution-focused one in the Gillian program analysis platform, which demonstrate the overall feasibility of our approach.
\end{enumerate}

\subsubsection*{Contribution 1: modular construction of CSE memory models.} 
Our key idea is a novel \emph{design pattern} that enables the construction of memory models for CSE tools using three concepts that we call \emph{memory-model atoms}, \emph{memory-model combinators}, and \emph{engine combinators}. 

Firstly, memory-model atoms (MMAs) describe the most basic memory models, such as values that abide by various well-known ownership disciplines, including exclusive, read-only, and fractional ownership, as per the SL combinator literature~\cite{Dockins09,Jung18}. 

Secondly, memory-model combinators (MMCs) take one or more memory models as parameters and return a new memory model that combines their functionalities. These include, for example:
\begin{itemize}
\item the standard {\em sum} and {\em product} combinators;
\item $\freeablestm$, which enables explicit book-keeping of memory deallocation, which is essential for sound true bug-finding; and
\item a collection of partial maps (which are the most often used data structure for representing memory), featuring various branching and consolidation strategies, capturing the choices implemented in {VeriFast~\cite{verifast-powerful-sound-jacobs-2011}, Viper~\cite{viper-verification-infrastructure-muller-2016}, and Gillian real-world memory models~\cite{gillian-part-multilanguage-fragososantos-2020,gillian-part-ii-maksimovic-2021,compositional-symbolic-execution-loow-2024}}.
\end{itemize}

Lastly, with engine combinators (ECs), we show how one can use combinators to capture general CSE concepts and analyses as CSE memory models, such as:
\begin{itemize}
\item management of user-defined predicates, essential for expressive unbounded verification; and
\item bi-abductive reasoning in the style of Infer-Pulse~\cite{finding-real-bugs-le-2022},
\end{itemize}
thereby decoupling them from core CSE engines, where their implementation traditionally belongs.

\subsubsection*{Contribution 2: evaluation.} We evaluate CSE à la carte in the following two ways:
\begin{itemize}
\item we mechanise our approach in Rocq on top of the CSE mechanisation of  \citet{compositional-symbolic-execution-loow-2025} by defining MMAs, MMCs, and ECs, and proving all relevant soundness results; and
\item we implement our approach in OCaml for the Gillian program analysis platform, to show that it scales to state-of-the-art CSE tools.\footnote{The mechanisation uses inference rules to model the operational semantics of the CSE engine. As Rocq cannot extract executable code from such inference rules, we wrote the OCaml implementation by hand following the~mechanisation.}
\end{itemize}

These two implementations allow us to evaluate CSE à la carte both theoretically and empirically. For the theoretical evaluation, in addition to the above-mentioned soundness results, we take a selection of mechanised monolithic memory models~\cite{compositional-symbolic-execution-loow-2025}, re-implement them using combinators, and prove behavioural equivalence between the original and re-implemented versions.

For the empirical evaluation, we re-implement all Gillian memory models, including the real-world ones of C and JavaScript (JS), except that of Gillian-Rust~\cite{hybrid-approach-semiautomated-ayoun-2025}, {for which we provide a pen-and-paper sketch only (for full details, see~\Cref{sec:eval-emp})}.
In doing so, we illustrate how our approach can capture idiosyncrasies of real-world memory models~(JS), hybrid memory models containing both monolithic and combinator-based components~(C), and advanced ghost state constructions~(Rust). Finally, we re-run all of the relevant previous case studies of Gillian, including C and JS case studies~\cite{gillian-part-multilanguage-fragososantos-2020,gillian-part-ii-maksimovic-2021,compositional-symbolic-execution-loow-2024}, using the new combinator-based memory models, revealing no unexpected behaviours and yielding no significant difference in performance.

\subsubsection*{Inspiration from past work.}
CSE à la carte brings together two streams of research, both using the idea of SLs parametric on partial commutative monoids (PCMs), which originated from abstract SL~\cite{local-action-abstract-calcagno-2007} and is used in the Views framework~\cite{views-compositional-reasoning-dinsdale-young-2013} and Iris~\cite{Jung18}. On the CSE side, our starting point is the {\em parametric CSE theory} of \citet{compositional-symbolic-execution-loow-2025}, which builds on earlier work on the Gillian platform~\cite{gillian-part-multilanguage-fragososantos-2020,gillian-part-ii-maksimovic-2021,compositional-symbolic-execution-loow-2024}. This meta-theory provides an instructive blueprint for building CSE tools for different languages by instantiating a core parametric framework with different {memory models} (MMs), which are, in essence, extensions of PCMs with operations for manipulating memory and assertions for specifying the memory. While this work does describe how CSE engines can be instantiated with different MMs, it stops short of providing a modular way for memory-model~construction.

On the SL side, we take the idea of building PCMs using {\em combinators}~\cite{Dockins09} to the CSE setting. This approach is well known from Iris~\cite{Jung18}, which introduced a series of combinators that underpin its version of PCMs, called \emph{resource algebras} (RAs). 
While both RAs and MMs are adaptations of PCMs, MMs contain substantially more information. In particular, RAs only describe the state and how to compose it; operations over the state (load, store, etc.) are defined separately and are therefore not built using RA combinators (in Iris terminology, such operations are called ``frame-preserving updates''). MMs, on the other hand, are much more expressive and describe {\em all}  parts of a parametric CSE engine that are dependent on the memory model as well as its meta-theory; specifically: 
\begin{itemize}
\item a \emph{concrete memory model} (CMM), describing memory-dependent concrete state, language operations that update the state, and the definition and interpretation of resources for specifying the state (such as the points-to assertion from SL);
\item a \emph{symbolic memory model} (SMM), describing memory-dependent symbolic state and language operations updating that symbolic state, including not only operations like load and store, but also, e.g., interactions with function specifications (the so-called consume and produce operations, cf.~\Cref{sec:overview}) that are essential for compositional reasoning; 
and 
\item \emph{soundness definitions and proofs} of the SMM w.r.t.~the CMM, which allow one to modularly prove an entire CSE engine sound for a new SMM by only proving this new SMM sound.

\end{itemize}

\subsubsection*{Paper structure.} We give an overview of our approach in \Cref{sec:overview} and restate the main concepts of parametric CSE in \Cref{sec:background}. Next, we present our memory-model atoms and combinators~(\Cref{sec:zoo}), followed by engine combinators~(\Cref{sec:zoo-engine}). We then evaluate our approach~(\Cref{sec:eval}), discuss limitations and threats to validity~(\Cref{sec:limitations}), provide an outline of related work~(\Cref{sec:related-work}), and give concluding remarks~(\Cref{sec:concs}).

%% file: sections/overview.tex
\section{Overview}\label{sec:overview}

\newcommand{\absentstm}{\stmfont{Absent}}

We give an overview of CSE à la carte, providing a modular way of constructing CSE tools using memory-model atoms and combinators. We first give a brief background on CSE, introducing the architecture of the CSE engines we work with in this paper, known as consume-and-produce engines, and the theory of~\citet{compositional-symbolic-execution-loow-2025}.

\subsection{CSE in a Nutshell}

A CSE engine is a type of symbolic execution engine inspired by SL reasoning, typically backed by an SMT solver. 
CSE engines work with {\em partial} memories and, in the spirit of the frame rule of SL, the analysis results obtained using a given memory are preserved under memory extension.

We work specifically with CSE engines based on the consume-and-produce paradigm, implemented by tools such as VeriFast, Viper, and Gillian. Such engines connect symbolic execution for analysing symbolic state  with  SL assertions using two operations called \emph{consume} and \emph{produce}: intuitively, consume matches an assertion with part of the symbolic state and removes it; produce extends the symbolic state with the resource corresponding to an assertion. Among other benefits, these operations  enable the  reuse of SL function specifications. Specifically, if a function comes with an SL specification, the corresponding function call can be executed symbolically by consuming the pre-condition of the specification and producing its post-condition.

 There are two fundamental notions of soundness for a CSE engine: \emph{over-approximate (OX)} and \emph{under-approximate (UX)}. OX soundness states that the symbolic engine captures all behaviours of the analysed language, ensuring that verification does not miss any bugs. UX soundness, instead, states that the engine captures only real behaviours of the analysed language, ensuring that any bugs found are true bugs. When a CSE engine is \emph{both} OX- and UX-sound, it is called \emph{exact}.

\subsection{Parametric CSE in a Nutshell}

We work with the CSE theory by~\citet{compositional-symbolic-execution-loow-2025}, which has several advantages for our use case over other CSE theories~\cite{local-reasoning-presence-raad-2020,featherweight-verifast-jacobs-2015,formal-foundations-translational-dardinier-2025,cn-verifying-systems-pulte-2023}: it is parametric on the memory model of the analysed language; and it supports {\em both} verification based on traditional SL and bug-finding based~on~ISL.

The analysed language $L$ is parametric on a concrete memory model (CMM): an arbitrary CMM is denoted by $\cmm$  and the instantiated language by $L(\cmm)$. The  corresponding CSE engine $\sym{L}$ is parametric on a symbolic memory model (SMM): an arbitrary SMM is denoted by
$\smm$ and the instantiated CSE engine by $\sym{L}(\smm)$ (``symbolicness'' is represented by the ``hat'' symbol). 
Both types of memory models have  basic memory operations (e.g., load and store), which are called \emph{memory actions}. 
In addition, CMMs provide \emph{resources}, atoms of SL assertions (e.g.,~a points-to assertion), and the interpretation of such resources in terms of concrete memory, while the SMMs provide symbolic consume and produce operations for these resources. Finally, $L(\cmm)$ effectively adds assignment, control flow, and function calls on top of $M$, and analogously for~$\sym{L}(\smm)$.

This theory also formalises soundness for the memory models and instantiated languages. 
 To uniformly refer to OX and UX soundness, we introduce the notion of $m$-soundness for $m \in \{ \mOX, \mUX \}$. We write  $\smm \sound_m \cmm$ to denote that the symbolic memory $\smm$ is $m$-sound with respect to the underlying concrete memory $\cmm$, that is, 
 all actions are $m$-sound, as are the consume and produce operations for all resources w.r.t.~the resource interpretation.
It also provides a \emph{lifting of soundness results} from the memory models to  the instantiated languages, which tells us that
$
\smm \sound_m \cmm \implies \sym{L}(\smm) \sound_m L(\cmm)
$.

\subsection{CSE à la Carte}
\label{sec:beebop}

While the CSE theory of~\citet{compositional-symbolic-execution-loow-2025} allows one to instantiate CSE engines with different memory models, what is missing is a way to {easily construct those memory models} as well as various analyses over them. To address this gap, we create  CSE à la carte, introducing  \emph{memory-model atoms} (MMAs), \emph{memory-model combinators} (MMCs), and \emph{engine combinators} (ECs),
offering substantial flexibility while dramatically reducing code size and maintenance effort. 

We start by introducing three MMAs, which constitute the ``leaves'' of memory models built up using MMCs, and which capture well-known SL ownership disciplines~\cite{Jung18} for \emph{one atom of memory} (say, a memory cell) containing a value of type $\type$. Specifically:
\begin{itemize}
\item $\exclstm(\type)$ represents standard exclusive ownership;
\item $\agstm(\type)$ represents standard read-only shared ownership (called ``agreement'' in Iris~\cite{Jung18}); and
\item $\fracstm(\type)$ represents standard fractional ownership.
\end{itemize}

Each of these MMAs comes with specific actions. In particular, $\exclstm(\type)$ and $\fracstm(\type)$ allow one to read ($\alookup$) and write ($\amutate$) the contents of the atom, whereas $\agstm(\type)$ allows one to read atom contents ($\alookup$) but, expectedly, not to write them.

Next, we introduce several MMCs for modularly constructing larger memory models from one or more input memory models. 
MMCs, in principle, lift constituent actions (with potential adjustments) and may introduce new actions of their own.
Our examples include:
\begin{itemize}
\item the sum ($M_1 \sumstm M_2$) and the product ($M_1 \prodstm M_2$) combinators, with their expected algebraic meaning, which lift constituent actions without changes and do not introduce new actions;
\item $\freeableoxstm(M)$ and $\freeablestm(M)$, which enable support for OX and exact deallocation, respectively, by extending $M$ with appropriate {\em deallocation} actions;%
\item several $\pmapstm(\tau, M)$ combinators, representing partial maps from keys of type $\tau$ to elements of memory model~$M$, which wrap the actions of $M$ into a map lookup, introduce an {\em allocation} action for creating a new key in the map (which, unlike deallocation, does not need a combinator of its own, since allocation is the same in OX, UX, and exact reasoning), and implement different lookup branching strategies;
\item $\dynmapstm(\tau, M)$, a partial map similar to $\pmapstm$ but in which keys can also be removed and in which the absence of keys can be tracked in line with the requirements of dynamically typed languages such as JavaScript. 
\end{itemize}

We also present two {\em engine} combinators, which formally do not differ in nature from MMCs but capture features not traditionally thought of as part of a memory model, beginning to explore how combinators can be used to capture concepts and analyses normally hard-coded into CSE engines:
\begin{itemize}
\item $\predstm(\predset, \cmm)$, which extends $M$ with the ability to reason about user-defined predicates that are described by $\predset$, and which introduces the {$\mathsf{fold}$} and {$\mathsf{unfold}$} actions; and 
\item $\bistm(\cmm)$, which extends $\cmm$ with UX bi-abductive reasoning and which is based on the fixes-from-missing-resource-errors approach pioneered by \citet{javert-20-compositional-fragososantos-2019} for JaVerT 2.0 and later adopted by Gillian.
\end{itemize}

Soundness of combinators is also expressed modularly, in the sense that if we have memory models $M_1, \dots, M_n$ and an $n$-parameter combinator with concrete variant $C$ and symbolic variant $\sym C$, 
then that combinator is $m$-soundness-preserving iff
\[
\smm_1 \soundm \cmm_1 \land \dots \land \smm_n \soundm \cmm_n \implies \sym{C}(\smm_1, \dots, \smm_n) \soundm C(\cmm_1, \dots, \cmm_n).
\]
Our soundness proofs are mechanised in Rocq. In particular, all three MMAs are proven to be OX- and UX-sound, and all combinators are proven to preserve their expected $m$-soundnesses: the $\freeableoxstm(M)$ combinator and the folding action of $\predstm(\predset, \cmm)$ are only OX-sound, as expected; all other actions and consume and produce operations are both OX- and UX-sound.

\subsection{Expressivity of CSE à la Carte}
\label{sec:simplythebest}

\subsubsection*{Basics: SL linear memory.} 
MMAs and MMCs allow for the construction of a broad range of memory models and associated soundness proofs through simple combinator composition. We start by capturing the demonstrator linear memory model of SL, with memory type $\nats \fparrow \val$,~as:
\[
\shade{\predstm(\predset, }\pmapstm(\nats, \freeableoxstm(\exclstm(\val)))\shade{)}.
\]
Whereas the use of $\pmapstm$ is self-evident, $\exclstm(\val)$ tells us that addresses contain exclusively owned values (cf.~\Cref{sec:zoo-own}), and $\freeableoxstm$ on top tells us that addresses can be deallocated (cf.~\Cref{sec:freeable}). Importantly, to enable unbounded verification, this and any other memory model should be additionally wrapped with a $\predstm$ combinator, shaded in \shade{\text{grey}} above.

\subsubsection*{Real-world: JavaScript.}
We now move to the memory model of JavaScript used for verification of real-world code, as per the work on JaVerT 2.0~\cite{javert-20-compositional-fragososantos-2019} and Gillian-JS~\cite{gillian-part-multilanguage-fragososantos-2020,gillian-part-ii-maksimovic-2021}. 
It is object-oriented and includes, first of all, a {\em heap} containing dynamic objects, each of which maps property names (strings) to complex JavaScript values, which we abstract here to $\val$. For soundness reasons,
each object also comes with an optional {\em domain}, which allows reasoning about absent object properties. Finally, each object has {\em metadata}, containing  the internal properties of that object, as prescribed by the  standard. The metadata was required for analysis tractability~\cite{cosette}, and is presented as a read-only resource. This model is straightforwardly constructed in our approach as:
\[
\pmapstm(\locs, \dynmapstm(\strs, \exclstm(\vals)) \prodstm \agstm(\locs))
\]
where $\locs$ denotes the type of object locations. Again, the top-level $\pmapstm$ is expected, mapping object locations to objects, which are represented as a product of two components. The first component, $\dynmapstm(\strs, \exclstm(\vals))$, captures both the heap and the domain as its dynamicity allows for the required reflection on and manipulation of absent object properties (cf.~\Cref{sec:mm-map-dom}). The second component, $\agstm(\locs)$, holds the address of the associated metadata object, and is declared shared and immutable (cf.~\Cref{sec:zoo-own}). This combinator-based model requires only 302 LoC to implement in Gillian, which is a reduction of 67\% when compared to the 919 LoC of the monolithic model, exhibits comparable performance, and comes with full soundness guarantees (cf.~\Cref{sec:eval-emp}).

\subsubsection*{Various flavours of partial maps.} 
Existing CSE tools commonly use partial maps to represent parts of their symbolic memories. 
However, when it comes to symbolic map operations, different tools have different branching strategies, consolidation strategies, and optimisations for efficiency. We believe that our combinator approach is flexible enough to encode practically any such strategy and optimisation. 
This we evidence by providing partial maps that perform: {eager vs.~lazy key equality resolution; SAT-based vs.~entailment-based branching}; and arbitrarily timed memory consolidation.

\subsubsection*{Custom combinators: register machines.}
Next, to illustrate custom combinators and how easy they are to create, let us capture an assembly-level memory model of a register machine comprising a linear heap and 32 machine registers. The linear heap component, $\mathit{LHeap}$, is defined exactly as above. 
When it comes to the registers, as a first try we could take $\textit{Regs} = \pmapstm([0, 32), \exclstm(\val))$, but this would allow us to create new registers (which is not permitted), as allocation is a built-in action of $\pmapstm$. The solution is to create a custom combinator (called, for example, $\mathit{PMap_{NoAlloc}}$), by taking the $\pmapstm$ combinator and filtering out its allocation action, yielding: $$\mathit{Regs} = \mathit{PMap_{NoAlloc}}(\pmapstm([0, 32), \exclstm(\val))).$$

Further, if we now tried to join these two components with the $\prodstm$ combinator, we could not capture the loading/storing of memory values into/from registers, as $\prodstm$ only lifts the actions of each component to the product but does not account for component interaction (see \appref{app:zoo}{A}). This, however, is not an issue, as we can extend the product combinator with these load-from-memory and store-from-register actions (calling this combinator, say, $\prodstm_\stmfont{RM}$), arriving at the solution:~\mbox{$\textit{LHeap} \prodstm_\stmfont{RM} \textit{Regs}$}.

Importantly, note how the proof burden in this case is limited only to the additional actions, as all previously existing actions are unaffected and have already been proven sound. 

\vspace*{-0.1cm}
\subsection{Implementations of CSE à la Carte}

We provide two implementations of our approach: one in Rocq on top of the mechanisation of \citet{compositional-symbolic-execution-loow-2025}; and one in OCaml for the Gillian engine. The Rocq implementation includes concrete and symbolic definitions as well as soundness proofs. The OCaml implementation only includes symbolic definitions, as the concrete definitions are only needed for the meta-theory.
In both implementations, memory models are represented as modules and~combinators~are~represented~as~functors.

Importantly, in both implementations, building stacks is as easy as on paper. For example, the following is the Rocq code (the OCaml code is similar) for instantiating the above linear memory model, where {\small \texttt{emb\_val\_in\_val}} and {\small \texttt{emb\_inf\_nat\_in\_val}} are modules used to embed values and natural numbers, respectively, inside values:
\begin{lstlisting}[language=Rocq]
Module linear_heap_ox := pmap emb_inf_nat_in_val (freeable_ox (ex emb_val_in_val)).
\end{lstlisting}

%% file: sections/background.tex

\section{Background: Parametric Compositional Symbolic Execution}%
\label{sec:background}

CSE à la carte builds on the parametric CSE theory of~\citet{compositional-symbolic-execution-loow-2025}. The theory is presented using a demonstrator language $\mathit{Lang}$ with a concrete and symbolic semantics both parametric on the \emph{memory model}. The language features a special command for performing \emph{memory actions}, as defined by the memory model, and a function-call command that, when executed symbolically, applies function specifications parametric on a separation logic whose atoms, called \emph{resources}, are likewise determined by the memory model.
To make this paper self-contained, we summarise the three central definitions of the theory: concrete memory models~(\Cref{sec:background-cmm}), symbolic memory models~(\Cref{sec:background-smm}), and OX and UX soundness (\Cref{sec:background-sound}). We slightly adapt the original definitions for consistency and omit user-defined predicates as these can now be handled by combinators (cf.~\Cref{sec:zoo-predicates}). We expect that most of the concepts used should be familiar to those who know parametric SLs, such as abstract SL~\cite{local-action-abstract-calcagno-2007}, Views~\cite{views-compositional-reasoning-dinsdale-young-2013}, and Iris~\cite{Jung18}. 

\subsection{Concrete Memory Models}%
\label{sec:background-cmm}

\begin{definition}\label{def:cmm}
A \emph{concrete memory model} is a tuple $\cmm = (\cmemss, \cwf, \emptycmem, \memcomp, \actions, \baction, \resources, \resmodels)$, where
\begin{itemize}
\item $\cmemss \ni \hp$ is the carrier type of concrete memories;

\item $\cwf \subseteq \cmemss$ is a well-formedness predicate;

\item $\emptycmem \in \cmemss$ is the (well-formed) empty memory;

\item $\memcomp \in \cmemss~\times~\cmemss \rightharpoonup \cmemss$ is a (well-formedness-preserving) partial composition operator such that $(\cmemss, \emptycmem, \memcomp)$ forms a partial commutative monoid (PCM);

\item $\actions \ni \action$ is the set of  memory actions;

\item $\hp.\action(\vec{v}) \baction \outcome: (\vec{v}', \hp')$ is the relation describing the action semantics, where 
$\vec{v}, \vec{v}' \in \vec{\val}$ are lists of $\mathit{Lang}$ values (e.g., Booleans, numbers, strings, etc.), and an outcome, $\outcome ::= \oxok \mid \oxerr \mid \oxm$, denotes either successful execution, a language error, or a missing-resource error\footnote{By explicitly modelling erroneous outcomes, we give a complete account of the behaviour of $\mathit{Lang}$ commands, not allowing the semantics to get stuck on any input. Missing-resource errors are errors that can be resolved by adding more memory, and must be singled out in various places (e.g., in the frame properties); they are also useful for powering bi-abduction~(\Cref{sec:zoo-biabduction}).}; this semantics must preserve the standard SL and ISL semantic frame properties\footnote{These standard frame properties describe how analysis results are maintained across larger memories and can be found in~\cite{compositional-symbolic-execution-loow-2025}; see also \citet{semantic-basis-local-yang-2002} for an accessible introduction to the SL frame property.} w.r.t.~composition;

\item $\resources \ni \resource$ is the set of resources; and

\item $\resmodels$ is a resource satisfaction relation, written $\hp \resmodels \resource(\vin; \vout)$, where  $\vin, \vout$ are lists of values representing the in- and out-parameters of the resource, with the idea that if one knows the in-parameters, one can learn the out-parameters (for more intuition on this, see the example of the linear memory model below and the discussion on symbolic consumption for the linear memory model in~\Cref{sec:background-smm}; further details can be found in~\cite{matching-plans-frame-loow-2024}).
\end{itemize}

\end{definition}

\begin{example}
The CMM of the linear memory model $\cmmlin$ is the tuple \[
(\cmemsslin, \cmemsslin, \emptyset, \uplus, \{ \alookup, \amutate, \anew, \afree \}, \baction^{\linsubscript}, \{ \mapsto \}, \resmodels^{\linsubscript})
\]
where: $\cmemsslin = \nats \fparrow \vals$; all memories are well-formed by construction; the empty memory is the empty map; composition is disjoint union of partial maps; there are four actions enabling value lookup and mutation and address allocation and deallocation; and there is one resource, the points-to resource~$\mapsto$, representing standard exclusive ownership of a memory address. To illustrate how memory actions work, consider (some of) the rules of the $\alookup$ action:
\begin{mathpar}\small
\infer{
  \hp.\alookup([n]) \baction^{\linsubscript} \oxok: ([v], \hp)
}{
  \hp(n) = v
}%
\and%
\infer{
  \hp.\alookup([n]) \baction^{\linsubscript} \oxm: ([\cmisscellmsg, n], \hp)
}{
  n \notin \dom(\hp)
}
\end{mathpar}
The points-to resource $\mapsto(n; v)$ (also written $n \mapsto v$) states that a memory cell at address $n$ holds value $v$. Note how $n$ is declared as an in-parameter and $v$ as an out-parameter because if we know $n$ we can learn $v$ by looking it up in memory. 
 The resource satisfaction relation $\resmodels$ is simply:
\[
\hp  \resmodels^{\linsubscript}
 n \mapsto v \iff \hp = \{n \mapsto v\}.
\]
\end{example}

\subsubsection*{Sketch of derived concrete definitions} Parametric CSE lifts CMMs to a concrete semantics of $\mathit{Lang}$,
which has: program variables, $\pvar x, \pvar y, \pvar z, \ldots \in \pvars$; program expressions, $\pexp \in \pexps$, which include standard logical and arithmetic operators; variable stores for evaluating expressions which do not affect the heap; and commands, $\cmd  \in \cmds$, which include simple commands, such as variable assignment, if-else conditional, sequential composition, function call, and also action call, $\passign{\vec{\pvar x}}{\action(\vec{\pexp})}$ with $\action \in \actions$.
The concrete semantics of $\mathit{Lang}$ is given by the judgement $\csesemtransabstract{\st}{\scmd}{\st'}{\fictx}{\outcome}$, where $\st, \st'$ are concrete states comprising a variable store and a concrete memory, $\scmd$ is a command, $\gamma$ is a set of function definitions provided as context to execute function calls, and $\outcome$ is the outcome of executing the command, the same one used in the action semantics.

CMMs also induce the standard assertions of SL and ISL. In particular, on top of logical expressions, $\lexp \in \LExp$, which extend program expressions with logical variables $x, y, z, \ldots \in \lvars$, the CMM resources induce the assertion language, $A \in \Assert^\resources::= \lexp \mid A \lstar A \mid \resource(\lexpin; \lexpout) \mid \dots$, with resource assertions
$\resource(\lexpin; \lexpout)$ for $\resource \in \resources$.
The satisfaction relation for assertions, $\subst, \st \asrtmodels A$, where $\subst$ is a substitution mapping logical variables to values, is induced by $\resmodels$ for $\resource$ and $\memcomp$ for $\lstar$.

\subsection{Symbolic Memory Models}%
\label{sec:background-smm}

Symbolic memory models (SMMs) require symbolic versions of the memory and actions,
as well as the \emph{consume} and \emph{produce} operations, which parametrically connect symbolic memory to resource assertions (used, e.g., for function specifications). SMMs are interpreted w.r.t.~an underlying CMM.

\begin{definition}\label{def:smm}
A \emph{symbolic memory model for a concrete memory model $\cmm$} (when obvious from context, we usually omit $\cmm$) is a tuple $\smm = (\smemss, \emptysmem, \memmodels, \baction, \resconsume, \resproduce)$, where:\footnote{The theory also includes oracles for standard angelic and demonic non-determinism, which are present in the mechanisation but elided from this presentation to avoid clutter, as they do not introduce anything novel w.r.t.~previous literature. In the few instances where we need to make an angelic choice, we simply write in the rules that we ``angelically pick''.}
\begin{itemize}
  \item $\smemss \ni \smem$ is the carrier type of symbolic memories;
  \item $\emptysmem \in \smemss$ is the empty symbolic memory;
  \item $\subst, \cmem^M \memmodels \smem$ is a satisfaction relation for symbolic memories, where $\subst$ is a substitution mapping logical variables to  values, with $\cmem^M \in \cmemss^\cmm$ and $\smem \in \smemss$;
  \item $\smem.\action(\vec{\sexp}, \spc) \baction_m o : (\vec{\sexp}', \smem', \spc')$ is the symbolic memory action relation, where $\action \in \actions^\cmm$ for memory action set $\actions^\cmm$,  $\vec{\sexp}, \vec{\sexp}' \in \vec{\SExp}$ are lists of symbolic expressions (i.e.,~logical expressions with no program variables), $\spc, \spc' \in \SExp$ are Boolean path conditions, $m \in \{ \mOX, \mUX \}$ is an approximation mode (over- or under-approximation),
  and $o ::= \oxok \mid \oxerr \mid \oxm \mid \oxabort$ is an outcome, where $\oxabort$ denotes an explicit incompleteness in the symbolic execution~engine;
  \item $\smem.\resconsume(\resource, \sexpin, \spc) \baction_m \outcome:  (\sexpout, \smem', \spc')$ is the consume operation, which removes (consumes) the resource $\resource \in \resources^\cmm$ with in-parameters $\sexpin$ in mode $m$ from the symbolic memory $\smem$ given path condition $\spc$, yielding an outcome $\outcome ::= \oxok \mid \oxabort$, out-parameters $\sexpout$, a new symbolic memory $\smem'$ in which $\resource(\sexpin; \sexpout)$ has been consumed, and a new path condition $\spc'$;
  \item $\smem.\resproduce(\resource, \sexpin, \sexpout, \spc) \baction (\smem', \spc')$ is the produce operation, which adds resource $\resource(\sexpin; \sexpout)$ to $\smem$ given path condition $\spc$, yielding a new symbolic memory $\smem'$ and a new path condition $\spc'$.
\end{itemize}
\end{definition}

\begin{example}
The linear symbolic memory model $\smmlin$ is the tuple:\[
(\smemsslin, \emptyset, \memmodels^{\linsubscript}, \baction_m^{\linsubscript}, \resconsume^{\linsubscript}, \resproduce^{\linsubscript})
\]
where  $\smemsslin = \SExp \fparrow \SExp$, the empty symbolic memory is the empty map, and the satisfaction relation is defined as a point-wise interpretation. Below, we provide this relation together with the successful cases of the $\alookup$ action and the consume and produce operations, whose resource argument is always $\mapsto$, the only resource of the linear memory model:
\begin{mathpar}\small
\begin{minipage}{0.45\textwidth}
\vspace*{-0.4cm}
$\subst, \hp \memmodels^\linsubscript \{ \sexp_a^1 \mapsto \sexp_e^1, \dots, \sexp_a^n \mapsto \sexp_e^n \}$ \\
$\hspace*{0.5cm} \iff \cmem = \uplus_{i = 1}^{n} \{ \esem{\sexp_a^i}{\subst} \mapsto \esem{\sexp_e^i}{\subst} \}$
\end{minipage}%
\and%
\infer{
  \smem.\alookup([\sexp_a], \spc) \baction_m^{\linsubscript} \oxok : ([\sexp_v],\smem, \spc')
}{
  \smem = \smem' \uplus \{ \sexp_a' \mapsto \sexp_v \} \quad \spc' = \spc \land (\sexp_a = \sexp_a')
}%
\\%
\infer{
  \smem.\resconsume^{\linsubscript}(\mapsto, [\sexp_a], \spc) \baction^{\linsubscript} \oxok : ([\sexp_v],\smem', \spc')
}{
  \smem = \smem' \uplus \{ \sexp_a' \mapsto \sexp_v \} \quad \spc' = \spc \land (\sexp_a = \sexp_a')
}%
\and
\infer{
  \smem.\resproduce^{\linsubscript}(\mapsto, [\sexp_a], [\sexp_v], \spc) \baction^{\linsubscript} (\smem', \spc')
}{
  \smem' = \smem \uplus \{ \sexp_a \mapsto \sexp_v \} \quad \spc' = \spc \land (\sexp_a \notin \dom(\smem))
}%
\end{mathpar}
The lookup rule branches over addresses $\sexp_a'$ in the symbolic memory $\smem$ that could equal $\sexp_a$, returning the corresponding binding and extending the path condition with the equality $\sexp_a = \sexp_a'$. 
The consume rule, in addition, removes the binding from the map, returning $\smem'$ instead of $\smem$. It receives the in-parameters of the resource (here, the address), and returns the out-parameters (here, the value stored at that address). Finally, the produce rule receives both the in- and the out-parameters, adding the corresponding binding to the symbolic memory. It also records in the path condition that the address being produced must not already be in the symbolic memory, ensuring that the resulting memory is still satisfiable.
\end{example}

\subsubsection*{Sketch of derived symbolic definitions} In parametric CSE, the symbolic semantics of the demonstrator language is obtained by lifting the resource-level consume and produce operations, $\resconsume$ and $\resproduce$, to the full assertion language using the set of assertions, $\Assert^\resources$ (\Cref{sec:background-cmm}),  and the operations on memory actions to the set of commands, $\cmds$ (\Cref{sec:background-cmm}). The top-level symbolic semantic judgement has the form $\cseenginetrans{m}{\fsctx}{\sst}{\scmd}{\outcome}{\sst'}$, where $\sst, \sst'$ are symbolic states (comprising a symbolic variable store, a symbolic memory, and a path condition), $\cmd$ is a  command, $m$ is a mode, and $\Gamma$ is a set of function specifications from SL or ISL depending on the mode (either OX or UX). Using the consume and produce operations, the symbolic semantics is able to use function specifications from $\Gamma$ to replace function calls with their specifications, enabling scalable compositional reasoning.

\subsection{Soundness}%
\label{sec:background-sound}

The soundness statements of the theory connect a symbolic memory model $\smm$ with its concrete memory model $\cmm$; the full definitions are rather elaborate, so we restrict ourselves to a sketch here.

\begin{definition}[Sketch of soundness of a memory model]
{A symbolic memory model $\smm$ for a concrete memory model $\cmm$ is said to be $m$-sound for $m \in \{ \mOX, \mUX \}$}, denoted by $\smm \soundm \cmm$, when:
\begin{itemize}
\item the symbolic memory action relation is $m$-sound w.r.t.~the concrete memory action relation;
\item the consume and produce operations of $\smm$, the $\resconsume$ and $\resproduce$, are $m$-sound w.r.t.~the satisfaction relation for resources, the $\resmodels$, and the composition operator, the $\memcomp$, of $\cmm$.
\end{itemize}
\end{definition}

\subsubsection*{Sketch of soundness lifting} Given $\smm$ and $\cmm$ with $\smm \soundm \cmm$, the theory provides two lifting theorems allowing the soundness of $\smm$ to be transferred to the instantiated engine. After instantiation, in simplified form, the OX and UX theorems, respectively, are as follows:
\[
\begin{gathered}
\csesemtransabstract{\st}{\scmd}{\st'}{\fictx}{\outcome} \land \st \models \sst \land (\forall \sst', \outcome.~\cseenginetrans{\mOX}{\fsctx}{\sst}{\scmd}{\outcome}{\sst'} \implies \outcome \neq \oxabort) \implies \exists \sst'.\; \st' \models \sst' \land \cseenginetrans{\mOX}{\fsctx}{\sst}{\scmd}{\outcome}{\sst'} \\[0.4em]
\cseenginetrans{\mUX}{\fsctx}{\sst}{\scmd}{\outcome}{\sst'} \land \st' \models \sst' \land \outcome \neq \oxabort \implies \exists \st.\; \st \models \sst \land  \csesemtransabstract{\st}{\scmd}{\st'}{\fictx}{\outcome}
\end{gathered}
\]
Both theorems come with abort-related conditions: the OX one requires $\oxabort$ to not be reachable, and the UX theorem only considers non-aborting execution paths. In~\citet{compositional-symbolic-execution-loow-2025}, an additional condition on $\oxm$ outcomes was needed as soundness of $\oxm$ outcomes cannot be guaranteed in the presence of user-defined predicates. This condition has  now moved inside the combinator for user-defined predicates (\Cref{sec:zoo-predicates}), which converts potentially unsound $\oxm$ outcomes into $\oxabort$~outcomes.

\subsubsection*{Connection with whole-program CMMs} 
Note that our CMMs are designed for CSE and therefore differ from more commonly used whole-program CMMs in that they are inherently partial (i.e.,~they allow for resource to be framed on and off) and may contain ghost state (e.g.,~fractional permissions or deallocation markers). Relating our CMMs to such whole-program CMMs is straightforward, and can be done using a standard simulation argument, as illustrated by \citet{compositional-symbolic-execution-loow-2025}. 

%% file: sections/zoo.tex

\section{Memory-model Atoms and Combinators}%
\label{sec:zoo}

We now introduce memory-model atoms and combinators. Our combinators construct complex memory models from simpler ones, automatically lifting memory actions, consume and produce operations, and all other components of memory models in such a way that soundness and additional properties, including frame-preservation, for the complex memory models hold by construction. When it comes to combinators, we focus on the most interesting ones, such as eager and lazy partial maps and deallocation, giving details of simpler ones in~\appref{app:zoo}{A}. In addition, we highlight the various insights that arose during the combinator creation and mechanisation~process.

\subsection{Memory-model Atoms}
\label{sec:zoo-own}

We introduce three MMAs, each capturing a specific \emph{ownership discipline} for a value~$v$ of type~$\type$: $\exclstm(\type)$ for exclusive (read-write), $\agstm(\type)$ for agreement (read-only), and $\fracstm(\type)$ for fractional (share xor write) ownership. We provide details on $\exclstm(\type)$ and $\agstm(\type)$ in \Cref{tab:atoms} and discuss~$\fracstm(\type)$ afterwards. We write $\type? \ni v_?$ to denote an option type: either $\Some(v)$ (eliding $\Some$ if clear from context) or~$\None$.

\begin{table}[t]
\caption{Excerpts of the definitions of the memory-model atoms $\exclstm(\type)$ and $\agstm(\type)$.}
\label{tab:atoms}
\small
\centering
\renewcommand{\arraystretch}{1.1}
\begin{tabular}{c|c|c}
 & {\bfseries Exclusive}: $\exclstm(\type)$ & {\bfseries Agreement}: $\agstm(\type)$  \\\hline\hline
$\cmemss$        & $\type?$ & $\type?$ \\\hline\hline
$\cwf$        & $\type?$ & $\type?$ \\\hline\hline
$\emptycmem$  & $\None$  & $\None$  \\\hline\hline
\multirow{2}{*}{$\memcomp$} & \multirow{2}{*}{$v_? \cdot \None = \None \cdot v_? = v_?$} & $v_? \cdot \None = \None \cdot v_? = v_?$ \\
& & $v \cdot v = v$  \\\hline\hline
$\actions$ & $\{ \alookup, \amutate \}$ & $\{ \alookup \}$ \\\hline\hline
\multirow{4}{*}{$\baction$} & $\None.\alookup([]) \baction \oxm : ([], \None)$ & $\None.\alookup([]) \baction \oxm : ([], \None)$ \\
& $v.\alookup([]) \baction \oxok : ([v], v)$ & $v.\alookup([]) \baction \oxok : ([v], v)$  \\\cline{2-3}
& $\None.\amutate([\_]) \baction \oxm : ([], \None)$ & \\
& $v.\amutate([v']) \baction \oxok : ([], v')$ & \\\hline\hline
$\resources$ & $\exclpred(; v)$ & $\agpred(; v)$ \\\hline\hline
$\resmodels$ & $v \resmodels \exclpred(; v)$ & $v \resmodels \agpred(; v)$ \\\hline\hline
$\resconsume$ & $\sexp.\resconsume(\exclpred, [], \spc) \baction \oxok : ([\sexp], \bot, \spc)$ & $\sexp.\resconsume(\agpred, [], \spc) \baction \oxok : ([\sexp], \sexp, \spc)$ \\\hline\hline
\multirow{2}{*}{$\resproduce$} & \multirow{2}{*}{$\None.\resproduce(\exclpred, [], [\sexp], \spc) \baction (\sexp, \spc)$} & $\None.\resproduce(\agpred, [], [\sexp], \spc) \baction (\sexp, \spc)$ \\
& & $\sexp.\resproduce(\agpred, [], [\sexp'], \spc) \baction (\sexp, \spc \land \sexp = \sexp')$
\end{tabular}
\end{table}

Both $\exclstm(\type)$ and $\agstm(\type)$ have as their carrier $\type?$ (with $\Some$ denoting ownership and $\None$ no ownership), have all memories be well-formed, and have $\None$ as the empty memory. Besides the standard composition with $\bot$, $\agstm(\type)$ also allows self-composition, making no memory exclusively owned. 

These two MMAs share the $\alookup$ action, whereas $\exclstm(\type)$ also has a $\amutate$ action; both actions return a missing outcome when called on $\bot$.
They also both have a single resource modelled by an appropriate non-empty memory. 

The symbolic variants of these two MMAs are constructed using a straightforward lifting from~$\type?$ to $\SExp?$. 
Consumption and production, the non-aborting cases of which are also given in \Cref{tab:atoms}, are as expected for $\exclstm(\type)$. 
Their behaviour for $\agstm(\type)$, however, to the best of our knowledge, is novel in the CSE setting:
note how successful consumption does not actually consume the (read-only) memory, and how production in a non-empty memory may also succeed, 
but only if the value in memory and the produced value are the same, captured by adding $\sexp = \sexp'$ to the path condition.

\subsubsection*{Fractional MMA} The $\fracstm(\type)$ MMA captures fractional ownership~\cite{fractional-permissions-boyland-2013}, where one can own a fraction $0 < q \leq 1$ of a resource: anyone who owns any fraction can read the value, but only someone who owns the full resource ($q = 1$) {can} modify it. This MMA is a straightforward generalisation of $\exclstm(\type)$. For example: the carrier is $(\type \times \mathbb{Q}_{(0, 1]})?$; the only resource, $\fracpred(q;v)$, has the fraction as an in-parameter, as otherwise one cannot know how much ownership to consume or produce; composition and production succeed only when the fraction sum does not overflow~1; consumption succeeds only if there is enough ownership to be subtracted; and so on. 

\subsubsection*{Relationship to Iris RAs}
Each of the three MMAs corresponds to an RA in Iris with the same name. As discussed in the introduction, our MMAs carry substantial additional information w.r.t.~Iris RAs. 

When it comes to $\exclstm(\type)$, the corresponding Iris RA also allows for OX-sound deallocation. However, as our MMAs are designed to be sound for both OX and UX analyses, we choose instead to factor out OX-sound and exact deallocation as part of the Freeable combinator~(cf.~\Cref{sec:freeable}).

On the other hand, the Iris $\mathit{Frac}$ RA only contains the fraction part of our pairs. For CSE, decomposing the fractional memory model further is not a good fit because of the need to define actions and consume and produce operations that manipulate the value and the fraction together.

\subsection{Eager Map}%
\label{sec:mm-map}

\newenvironment{insight}
    {\par\addvspace{\topsep}\leftskip=1em}
    {\par\addvspace{\topsep}}

Finite partial maps are ubiquitous in SL reasoning. They are used to model physical and ghost heaps, objects, stacks, and registers, and we use them in \emph{every instantiation}. Implementing them as combinators reduces codebase size (e.g., 17\% for a single partial map in Gillian-C) and factors out performance- and soundness-sensitive code. In the context of symbolic execution, maps can be implemented in different ways, with different trade-offs, including the choice of {\em branching strategies}. We show that our combinator approach is flexible enough to capture all strategies employed in modern CSE tools, and do so starting from what we call ``eager'' maps.

\subsubsection*{Concrete MM}
The concrete eager finite partial map combinator is denoted by $\pmapstm(\type, \cmm)$, where $\type$ denotes the key (domain) type and $M$ is an arbitrary memory model denoting the value (codomain) type. A memory is then, expectedly, a map from keys to values in $\cmemss^\cmm$: $\cmemss \defeq \type \fparrow \cmemss^\cmm$. It is well-formed if all values are well-formed (according to $\cwf^\cmm$) and non-empty (different from~$\emptycmem^\cmm$).
The empty memory is the empty map, $\emptycmem \defeq \emptyset$, and composition is defined~pointwise:
\[\small
(\cmem_1 \memcomp \cmem_2)(v) \defeq
\begin{cases}
\cmem_1(v), \text{if } v \in \dom(\cmem_1) \setminus \dom(\cmem_2); & \quad\cmem_1(v)\memcomp^\cmm \cmem_2(v), \text{if } v \in \dom(\cmem_1) \cap \dom(\cmem_2);  \\
\cmem_2(v), \text{if } v \in \dom(\cmem_2) \setminus \dom(\cmem_1 ); & \quad\text{undefined}, \text{otherwise}. \\
\end{cases}
\]

\begin{insight}
\noindent{\bfseries Insight: Memory-like PCMs.}
Much like~\citet{Dockins09}, who restrict their considered PCMs to build a more ``pleasing'' theory, we also restrict our PCMs to avoid edge cases that do not occur in memories in practice and would only serve to clutter the definitions with side conditions. Specifically, we require PCMs that are inputs to combinators to be non-empty and to not have memories that cancel out (that is, $\cmem \neq \emptycmem \implies \cmem \memcomp \cmem' \neq \emptycmem$, a condition also known as {\em positivity}). 

\end{insight}

The set of actions of $\pmapstm(\type, \cmm)$ is $\actions \defeq \{ \asub(\action) \mid \action \in \actions^\cmm \} \cup \{ \anew \}$, that is, all actions from $\cmm$ are lifted and tagged with $\asub$, and there is an additional $\anew$ action that adds a fresh binding to the map. Their semantics is as follows:
\begin{mathpar}%
\small%
\inferrule{
  \mu(k).\action(\vec{v}) \baction \outcome: (\vec v', \cmem^\cmm) \\\\
  \mu' = \mu \setminus \{ k \}    \text{ if } \cmem^\cmm = \emptycmem^\cmm \text{ else } \mu[k \mapsto \cmem^\cmm]
}{
  \mu.\asub(\action)(k \cons \vec{v}) \baction \outcome: (\vec v', \mu')
}
\and
\inferrule{
  k \notin \dom(\mu) \quad
  \mu' = \mu \uplus \{ k \mapsto \cmm.\initcmem \}
}{
  \mu.\anew([]) \baction \oxok: ([k], \mu')
}
\end{mathpar}
The semantics of lifted actions $\asub(\action)$ is mostly straightforward: they take an additional key argument, identify the corresponding cell in the map, and execute $\action$ on that cell, leaving the rest unchanged. If $\action$ results in $\emptycmem^\cmm$, the key is removed from the map to maintain well-formedness.

\noindent
The $\anew$ action receives no arguments and allocates a new cell in the map, returning a fresh key and the updated memory with the new cell initialised with a dedicated initial value $\cmm.\initcmem$ provided by $\cmm$. The value $\cmm.\initcmem$ is required to satisfy the following properties: it must be well-formed, $\cwf^\cmm(\cmm.\initcmem)$; it must not be empty, $\cmm.\initcmem \neq \emptycmem^\cmm$, to preserve map well-formedness; and it must be {\em exclusive} in UX mode (discussed below). The $\anew$ action gives us the following two insights.

\begin{insight}
\noindent{\bfseries Insight: Combinator-custom constants and operations.}
Some combinators introduce their own special constants and operations (for example, the $\initcmem$ of eager maps and the consolidation and symbolic composition of lazy maps introduced shortly), which must be lifted along with the actions and the consume and produce operations.
Such symbolic constants and operations must also be proven $m$-sound with respect to their concrete counterparts.
\end{insight}

\begin{insight}
\noindent{\bfseries Insight: Exclusive memories.}
Like RA combinators, some of our combinators require \emph{exclusive memories}. A memory $\cmem \in \cmm$ is \emph{exclusive} if it cannot be composed with any other memories, i.e.,~$\cmem \memcomp \cmem'$ is defined iff $\cmem' = \emptycmem$; in other words, it has no frames. For example, a non-empty $\cmem \in \exclstm(\type)$ is always exclusive, $\cmem \in \agstm(\type)$ never, and $\cmem \in \fracstm(\type)$ iff its fractional permission is 1.

As per the above rule for map allocation, the initial value $\cmm.\initcmem$ must be exclusive in $\mux$ mode, as otherwise allocation would break the UX frame property. This prohibits, for instance, $\agstm(\type)$ from being used in the codomain for UX analysis. As a consequence, the JavaScript memory model for OX verification as presented in \Cref{sec:overview} cannot be used as-is for UX bug-finding, but can be appropriately adjusted with a custom mode-specific read-only memory model (cf.~\Cref{sec:eval-emp}).
\end{insight}

The resources of the partial map combinator are $\resources \defeq \resources^\cmm$, but lifted per map key in that they receive an additional in-parameter denoting the key. The satisfaction relation thus becomes:
\[\cmem \resmodels \resource(k \cons \vin; \vout) \iff \exists \cmem'.~\cmem = \{ k \mapsto \cmem' \} \land \cmem' \resmodels^\cmm \resource(\vin; \vout) \land \cmem' \neq \emptycmem^\cmm\]

\begin{example}\label{ex:fractional}
The standard linear memory model with fractional ownership~\cite{fractional-permissions-boyland-2013} corresponds to $\pmapstm(\nats, \fracstm(\val))$. The action $\alookup([])$ of $\fracstm(\val)$ lifts to $\asub(\alookup)([k])$, which receives the key (or address) $k$ as an argument and returns the corresponding value. Similarly, the $q \mapsto v$ resource of the $\fracstm(\val)$ codomain is lifted to the standard fractional points-to resource,~$k \fmapsto{q} v$.
\end{example}

\subsubsection*{Symbolic MM}
The symbolic eager finite partial map combinator, $\spmapstm(\type, \smm)$, is defined similarly to the concrete one: symbolic memories are maps from symbolic expressions of type $\type$ to symbolic values: $\smemss \defeq \SExp \fparrow \smemss^{\smm}$; and the empty symbolic memory is the empty map, $\emptysmem \defeq \emptyset$. 

The eager map is eager in that it resolves address equalities as early as possible. This is most easily seen in the definition of the satisfaction relation of symbolic memories, which enforces that syntactically different symbolic keys evaluate to different concrete keys:
\begin{equation}
\label{eq:memsat}
\small
\inferrule
 {}
 {\subst, \emptyset \memmodels \emptyset}
\qquad
 \inferrule
 {\sexp \notin \dom(\smem') \quad \subst, \cmem' \memmodels \smem'  \quad \esem{\sexp}{\subst} = v \shade{\notin \dom (\mu')} \quad \smem^{\smm} \neq \emptysmem^{\smm}  \quad  \subst, \cmem^M \memmodels^{\smm} \smem^{\smm} }
 {\subst,  \cmem' \cdot \{ v \mapsto \cmem^M \} \memmodels \smem' \uplus \{ \sexp \mapsto \smem^{\smm} \}}
\end{equation}
In particular, this means that $\subst, \cmem \memmodels \smem \implies |\dom(\cmem)| = |\dom(\smem)|$, that is, a symbolic map $\smem = \{ (\sexp_i \mapsto \smem^\cmm_i)|_{i=1}^n \}$ can only be satisfied by concrete maps with $n$ bindings, $\cmem = \{ (\esem{\sexp_i}{\subst} \mapsto \cmem_i^\cmm)|_{i=1}^n\}$, where $(\subst, \cmem_i^\cmm \memmodels^{\smm} \smem_i^\cmm)|_{i=1}^n$. This is not true for the lazy version of the map combinator, introduced in the next section, for which we remove the highlighted premise, $v \notin \dom (\mu')$.

Within eager maps, when it comes to subaction lifting, we formalise two branching strategies, a SAT-based and an entailment-based one:
\begin{mathpar}%
\small%
\inferrule[\text{Subaction-SAT}]{
  \shade{\sexp_a' \in \dom(\smem)} \quad \smem(\sexp_a') \neq \smem^{\smm}_\emptyset \quad
  \shade{\spc'' = \spc \land (\sexp_a = \sexp_a')} \\\\
  \smem(\sexp_a').\action^{\smm}(\sexps, \shade{\spc''}) \baction_m^{\smm} \outcome : (\sexps', \smem^{\smm}, {\spc'}) \\\\
  \smem' = \smem \setminus \{ \sexp_a' \} \text{ if } \smem^{\smm} = \emptysmem^{\smm} \text { else } \smem[\sexp_a' \mapsto \smem^{\smm}]
}{
  \smem.\asub(\action)(\sexp_a \cons \sexps, \spc) \baction_m^{\pmapstm(\type, \smm)} \outcome : (\sexps',\smem', {\spc'})
}
\and%
\inferrule[\text{Subaction-Entailed}]{
  \shade{\text{angelically pick }\sexp_a'} \quad
  \smem(\sexp_a') \neq \smem^{\smm}_\emptyset \quad
  \shade{\spc \models \sexp_a = \sexp_a'} \\\\
  \smem(\sexp_a').\action^{\smm}(\sexps, \shade{\spc}) \baction_m^{\smm} \outcome : (\sexps', \smem^{\smm}, {\spc'}) \\\\
  \smem' = \smem \setminus \{ \sexp_a' \} \text{ if } \smem^{\smm} = \emptysmem^{\smm} \text { else } \smem[\sexp_a' \mapsto \smem^{\smm}]
}{
  \smem.\asub(\action)(\sexp_a \cons \sexps, \spc) \baction_m^{\pmapstm(\type, \smm)} \outcome : (\sexps',\smem', {\spc'})
}
\end{mathpar}
Contrast the two strategies, with differences given in \shade{\text{grey boxes}} above: the SAT-based one branches on equality between the input key $\sexp_a$ and all existing keys in the memory $\sexp_a'$, whereas the entailment-based one discovers an address $\sexp_a'$ provably equal to the lookup address $\sexp_a$.\footnote{A common discovery method is to first check for syntactic equality via direct map lookup and, if that fails, iterate over map addresses until success or until all keys have been checked unsuccessfully. We abstract the discovery away via angelic non-determinism, showing that specific details do not matter for soundness as long as a working address is found.} 
Once the key $\sexp_a'$ has been identified, both strategies execute the corresponding codomain action on the corresponding cell, leaving the rest of the map unchanged. Note that, should the action result in the empty memory, the key $\sexp_a'$ is removed from the map as symbolic memories are not allowed to be empty. In addition, as a detail of proving UX soundness, we must require the looked-up memory, $\smem(\sexp_a')$, to be non-empty.

It is important to note that, while the SAT-based strategy appears to cause excessive branching, in practice almost all branches~are pruned away instantly, without reaching the SMT solver. On the other hand, while the entailment-based strategy branches less than the SAT-based one, it is not applicable in as many contexts, as per the following example.

\begin{example}\label{ex:demonic}
In the fractional permissions MM of \Cref{ex:fractional}, let $\smem = \{\sym{x} \mapsto (42, 0.5), \sym{y} \mapsto (42, 0.5)\}$ and let $\spc = \sym{z} \in \{ \sym{x}, \sym{y} \}$, and let us try to execute $\smem.\asub(\alookup)([\sym{z}], \spc)$. The SAT-based strategy simply branches over $\sym{z} = \sym{x}$ and $\sym{z} = \sym{y}$ and execution continues. In contrast, the entailment-based strategy does not apply at all as neither $\sym{z} = \sym{x}$ nor $\sym{z} = \sym{y}$ is provable from $\spc$, and execution stops.
\end{example}

Given the action rules, the consume and produce rules are relatively straightforward. We illustrate by giving two SAT-based rules for produce, one adding a new mapping (\textsc{Produce-New-SAT}) and one branching over existing addresses (\textsc{Produce-Update-SAT}), analogously to \textsc{Subaction-SAT}:
\begin{mathpar}
\small
\inferrule[\text{Produce-New-SAT}]{
  \spc' = \spc \land \sexp_a \notin \dom(\smem) \\\\
  \emptysmem^{\smm}.\resproduce^{\smm}(\resource, \sexpin, \sexpout, \spc') \baction^{\smm} (\smem^{\smm}, \spc'')
}{
  \smem.\resproduce(\resource, \sexp_a \cons \sexpin, \sexpout, \spc) \baction (\smem[\sexp_a \mapsto \smem^{\smm}], \spc'')
}
\and
\inferrule[\text{Produce-Update-SAT}]{
  \smem(\sexp_a') \neq \smem^{\smm}_\emptyset \quad
  \spc' = \spc \land (\sexp_a = \sexp_a') \\\\
  \smem(\sexp_a').\resproduce^{\smm}(\resource, \sexpin, \sexpout, \spc') \baction^{\smm} (\smem^{\smm}, \spc'')
}{
  \smem.\resproduce(\resource, \sexp_a \cons \sexpin, \sexpout, \spc) \baction (\smem[\sexp_a' \mapsto \smem^{\smm}], \spc'')
}
\end{mathpar}

\begin{example}\label{ex:eager}
Consider a function \textsf{f} with the specification $\triple{\pvar{x} \fmapsto{0.5} 42 \lstar \pvar{y} \fmapsto{0.5} 42}{\textsf{f}(\pvar{x}, \pvar{y})}{\dots}$, using again the fractional permissions MM presented in \Cref{ex:fractional}. In CSE tools, the verification of $\textsf{f}$ begins by producing its pre-condition in the empty symbolic memory, yielding symbolic states that represent it. This pre-condition contains two resources, $\fracpred(x, 0.5; 42)$ and $\fracpred(y, 0.5; 42)$, which will be produced in succession. The first resource is handled by \textsc{Produce-New-SAT}. When producing the second resource, however, both rules apply and production branches on whether $\sym{y} = \sym{x}$ (as address equality is resolved eagerly), resulting in the end in the \emph{two} following symbolic~states:
\[
\small
\begin{array}{c}
(\smem_1, \spc_1) = (\{ \sym{x} \mapsto (42, 0.5); \sym{y} \mapsto (42, 0.5) \}, \sym{x} \neq \sym{y}) \qquad
(\smem_2, \spc_2) = (\{ \sym{x} \mapsto(42, 1) \}, \sym{x} = \sym{y})
\end{array}
\]
\end{example}

\subsection{Lazy Map}%
\label{sec:consolidation}

\newcommand{\exsubscr}{\text{ex}}

We introduce a ``lazy'' version of the map combinator, $\pmapconsstm$, where address equality resolution can be delayed arbitrarily.
This approach captures the notion of \emph{consolidation}, used in Viper~\cite{advancing-automated-permissionbased-schwerhoff-2016,Eilers24}, JaVerT~\cite{javert-20-compositional-fragososantos-2019}, Gillian-JS, and Gillian-C~\cite{gillian-part-ii-maksimovic-2021} for reasoning about objects.
As the branching strategy only affects the symbolic combinator, the concrete combinator $\pmapconsstm$ is identical to $\pmapstm$. In what follows, we discuss the differences between the symbolic variants, $\spmapconsstm$ and $\spmapstm$.

\subsubsection*{Symbolic memory satisfaction relation and memory operations} As discussed earlier (cf.~\Cref{eq:memsat}), the only difference in the symbolic memory satisfaction relation between eager and lazy maps, but a critical one, is that the latter drops the address disjointness requirement, $v \notin \dom (\mu')$.

\begin{example}
Let $\smem = \{ \sym{x} \mapsto (42, 0.5); \sym{y} \mapsto (42, 0.5) \}$. 
For $\spmapstm(\nats, \sfracstm(\val))$, this memory only has one model (modulo choice of $\subst$), namely $\subst, \{ \esem{\sym{x}}{\subst} \mapsto (42, 0.5); \esem{\sym{y}}{\subst} \mapsto (42, 0.5) \} \memmodels \smem$ with $\esem{\sym{x}}{\subst} \neq \esem{\sym{y}}{\subst}$, as per the syntactic key separation of eager maps. 
For $\spmapconsstm(\nats, \sfracstm(\val))$, however, we also have the model $\subst, \{ \esem{\sym{x}}{\subst} \mapsto (42, 1) \} \memmodels \smem$ with $\esem{\sym{x}}{\subst} = \esem{\sym{y}}{\subst}$, as the two half-permission cells may actually represent a single full-permission cell as a result of delayed branching.
\end{example}

Thanks to the lazy satisfaction relation, operations over memory can now soundly defer branching, ultimately allowing syntactic address lookups, and the one that benefits the most from this is the produce operation.
In particular, the production rules of the lazy map rely only on syntactic address lookup: for example, the lazy SAT-based produce rules (compacted into one below) now consider $\smem(\sexp_a)$ directly and do not adjust the path condition:
\begin{mathpar}
\small
\inferrule[\text{Produce-SAT-Lazy}]{
  \smem^{\smm}_1 \eqdef \smem^{\smm}_\emptyset \text{ if } \sexp_a \notin \dom(\smem) \text{ else } \smem(\sexp_a) \quad 
  \sexp_a \in \dom(\smem) \implies \smem^{\smm}_1 \neq \smem^{\smm}_\emptyset \quad
  \smem^{\smm}_1.\resproduce^{\smm}(\resource, \sexpin, \sexpout, \spc) \baction^{\smm} (\smem^{\smm}_2, \spc')
}{
  \smem.\resproduce(\resource, \sexp_a \cons \sexpin, \sexpout, \spc) \baction (\smem[\sexp_a \mapsto \smem^{\smm}_2], \spc')
}
\end{mathpar}

\begin{example}\label{ex:frac-lazy}
Consider the same example as in \Cref{ex:eager}, but using the lazy fractional permissions MM, $\pmapconsstm(\nats, \fracstm(\val))$. Producing a binding for the key $\sym{y}$ when $\sym{x}$ is already present does not require branching, but simply adds a new binding for $\sym{y}$ to the map, even if $\sym{x}$ and $\sym{y}$ may evaluate to the same concrete key, resulting in a unique symbolic state:
\[
\small
\begin{array}{ll}
(\smem, \spc) = (\{ \sym{x} \mapsto (42, 0.5); \sym{y} \mapsto (42, 0.5) \}, \true)
\end{array}
\]
Because of the lazy satisfaction relation, this single symbolic state represents the same set of concrete states as the two symbolic states of the eager map (\Cref{ex:eager}).
\end{example}

\subsubsection*{Consolidation} 
Branching, however, cannot be delayed forever. 
To illustrate, we continue \Cref{ex:frac-lazy} and additionally assume that the body of \textsf{f} is $\pifelse{\pvar{x} = \pvar{y}}{\{~\amutate(\pvar{x}, 43)~\}}{\{~\dots~\}}$. 
In the then-branch, after learning that $\sym{x} = \sym{y}$, one ought to be able to mutate the cell at address $\sym{x}$, since the resource is owned with a fraction of $1$; yet the symbolic $\amutate$ action will fail, because the resource is scattered across two bindings. 
This is where \emph{consolidation}~\cite{advancing-automated-permissionbased-schwerhoff-2016} comes in: at any point after learning new equalities, lazy maps can be consolidated by merging bindings whose keys have been found equal. 
In a real CSE engine, consolidation can happen at any point during execution. 
In our Rocq mechanisation, we capture this by introducing a $\mathsf{consolidate}$ ghost command which can be invoked anywhere, leaving the choice of \emph{when} to insert this action during execution abstract. 
A consolidation action is $m$-sound when it is equivalent to a concrete no-op operation.

\begin{wrapfigure}{r}{0.37\textwidth}
\vspace{-0.65cm}
\hspace*{-0.34cm}
\begin{minipage}{0.410\textwidth}\centering
\begin{mathpar}\small
\inferrule{
  \text{angelically pick } \sexp_a, \sexp_b \quad \sexp_a \neq \sexp_b\\\\
  \spc \models \sexp_a = \sexp_b \quad \smem^{\smm}_\emptyset \notin \{ \smem(\sexp_a), \smem(\sexp_b) \} \\\\
  \smem(\sexp_a)\ \sym{\cdot}\ \smem(\sexp_b) = (\smem^{\smm}, \spc'') \quad \spc' = \spc \land \spc'' \\\\
  \smem' = (\smem \setminus \{ \sexp_b \})[\sexp_a \mapsto \smem^{\smm}]
}{
  \mathsf{consolidate}^{\pmapconsstm(\tau, \smm)}(\smem, \spc) \rightsquigarrow (\smem', \spc')
}%
\end{mathpar}
\end{minipage}
\vspace{-1em}
\end{wrapfigure}

For the lazy map, consolidation is formally defined by the rule to the right. To define it, we require a custom symbolic composition operator on the memory model the combinator is lifting: $\smem_1\ \sym{\cdot}\ \smem_2 = (\smem, \spc)$, which must be $m$-sound w.r.t. concrete composition. The extra path condition $\spc$ of symbolic composition describes the constraint under which composition is defined. For example, for memories in $\smemss^{\fracstm(\tau)}$, the composition $(\sexp_1, \sym{q}_1)\ \sym{\cdot}\ (\sexp_2, \sym{q}_2)$ yields $((\sexp_1, \sym{q}_1 + \sym{q}_2), \sexp_1 = \sexp_2 \land \sym{q}_1 + \sym{q}_2 \leq 1)$, requiring $\sexp_1$ and $\sexp_2$ to be the same and the sum of the fractions not to exceed~$1$. With symbolic composition in place, the idea of the consolidation rule is as follows: we discover two different addresses $\sexp_a$ and $\sexp_b$ that are equal under $\spc$ (again, we abstract   search specificities by angelic non-determinism); if the memories at these addresses can be symbolically composed, then the mapping for $\sexp_b$ is removed, and $\sexp_a$ is mapped to the merged~memory.

\subsubsection*{Soundness} We prove that the lazy map is both $\mox$- and $\mux$-sound with respect to the \emph{same concrete memory model} as the eager map. In other words, both symbolic map implementations are interchangeable and can be used for the same reasoning tasks, with the same guarantees.

\subsubsection*{Discussion} The choice between eager and lazy maps is a trade-off between performance and automation. Eager maps often require more branching, which may affect performance, but ensure that the engine always has a precise understanding of the memory. On the other hand, while consolidation helps, when to apply consolidation and what keys to merge depend on heuristics, and lazy maps may fail to identify the required resource. A tool may then require additional user annotations to ask the verification engine to perform this consolidation, hence affecting automation.

Note that a single memory model may even use both variants of the map at the same time.
In fact, the Gillian-JS memory model (cf.~\Cref{sec:eval-emp}) uses lazy maps from object locations to objects and eager maps from field names to field values within each object. Eagerly branching on object equality would lead to an explosion of branches, as any two objects with different fields could be equal, whereas the use of symbolic field names within objects is rare enough, and potentially more impactful on automation, that the use of an eager map is justified.

Finally, we note that there could be many more variations of maps with similar concrete semantics but different symbolic behaviour; we present eager and lazy, but see no limitations to implementing more. Our modular framework could allow tool developers to easily explore this design space and make informed decisions about the trade-offs involved in different contexts. Such experiments have been performed in the past by \citet{Eilers24}, who explored a similar kind of trade-off within Viper, but they ``had to re-implement the heap representation [and] all heap-manipulating operations'' for their experiments rather than just switching out one map combinator for another.

\subsection{Freeable}%
\label{sec:freeable}

We  model deallocation using 
the $\freeablestm$ combinator, which adds a deallocation action to a given memory model $\cmm$: $\actions \defeq \{ \asub(\action) \mid \action \in \actions^\cmm \} \cup \{ \afree \}$.

\subsubsection*{Freeable for OX reasoning}

For the OX variant, we have that $\freeableoxstm(\cmm) \defeq ``\cmm$ with an additional $\afree$ action'', with the interesting cases of the $\afree$ action semantics being as follows:
\begin{mathpar}\small
\infer{
 \hp.\afree([]) \baction \oxok: ([], \emptycmem^\cmm)
}{
 \hp \text{ is exclusive}
}
\and
\infer{
 \hp.\afree([]) \baction \oxm : ([], \hp)
}{
 \hp \text{ is not exclusive}
}
\end{mathpar}

The exclusivity requirement captures the well-known property that a cell can be deallocated {\em iff} fully owned: for $\pmapstm(\nats, \freeableoxstm(\exclstm(\val)))$, $\asub(\afree)([k])$ requires $k$ to be in the map; for  $\pmapstm(\nats, \freeableoxstm(\fracstm(\val)))$, $\asub(\afree)([k])$ requires $k$ to be in the map with fraction $1$. Note how $k$ is effectively removed from the $\pmapstm$ by updating its value to $\emptycmem^\cmm$ (cf.~action lifting in~\Cref{sec:mm-map}).

\subsubsection*{Freeable for exact reasoning}

As discussed in~\Cref{sec:zoo-own}, deallocated resources cannot simply be forgotten for UX and exact reasoning; they have to be explicitly tracked.
We capture this with $\freeablestm(\cmm) \defeq \cmm \sumstm \exclstm(\{ \freed \})$, where $\sumstm $ is the sum combinator (given in~\appref{app:zoo}{A}) and $\freed$ is a special element indicating that the memory has been freed, with, as above, an additional $\afree$ action. The most interesting action rules are, where $\inl$ and $\inr$ are, respectively, the left and right~constructors~of~sum: 
\begin{mathpar}\small
\and
\infer{
 \inl(\cmem).\afree([]) \baction \oxok: ([], \inr(\Some(\freed)))
}{
 \mu \text{ is exclusive}
}%
\and%
\infer{
 \cmem.\action(\vs) \baction \oxerr: ([``\textsf{use-after-free}"], \cmem)
}{
  \cmem = \inr(\Some(\freed)) \and \action \in \actions^\cmm
}
\end{mathpar}
from which we see, for example, that all actions on a freed memory yield a use-after-free error.

We know from Iris that an update that ``moves'' the result from one sum component to another, say from left to right (as does $\afree$), is OX frame-preserving only if the \emph{left} state $\inl(\cmem)$ is exclusive. In the UX setting, which is not covered by Iris, we observe that the \emph{right} state must also be exclusive (which $\afree$'s $\Some(\freed)$ indeed is). This, interestingly, means that some RAs frequently used in Iris, such as the $\mathit{OneShot}$ algebra (in our notation: $\exclstm(\{ \freed \}) \sumstm \agstm(\type)$)~\cite{Jung18}, are incompatible with UX reasoning, as no element of $\agstm(\type)$ can ever be exclusively owned. 

Lastly, lifting actions has a subtlety concerning {\em allocating actions}: actions that can succeed when executed concretely from $\emptycmem$ (e.g., the $\anew$ action of $\pmapstm$).

\begin{insight}
\noindent{\bfseries Insight: Allocating actions.}
Lifting allocating actions can break both frame properties, so they must not be lifted. In $\freeablestm(\cmm)$, an allocating action succeeds from $\emptycmem$ but, like every action, yields a use-after-free error on the freed cell $\inr(\Some(\freed))$, a frame of $\emptycmem$. The OX frame property fails, as the error on the freed cell is matched from $\emptycmem$ by neither a miss nor an error. The UX frame property fails for actions that can succeed from $\emptycmem$ and return the memory unchanged: their result $\emptycmem$ composes with the freed cell, so their success would have to be reproducible there. Separately, lifting non-allocating actions sometimes requires them to be total and deterministic w.r.t.~their outcome from $\emptycmem$, and to leave the memory empty on an error from $\emptycmem$.
\end{insight}

\subsection{Map-with-domain}%
\label{sec:mm-map-dom}

As demonstrated by the Gillian-JS memory model in \Cref{sec:overview}, partial maps can be used to model objects within the heap as well as the top-level heap.
To allocate and free such objects, we need to establish a way of making these maps \emph{exclusive} (cf.~\Cref{sec:zoo-own,sec:freeable}), ensuring that no frame, that is, an additional binding, may be added to them. Several tools address this by adding a resource characterising the ``domain'' of the map.
In CompCert-like memory models~\cite{formally-verified-compiler-leroy-2009}, where objects are arrays of bytes, the domain of each object is a non-negative integer denoting the allocated size of the object, asserting that only offsets below this bound exist. Such a bound resource exists in tools such as VeriFast, Gillian-C, and RefinedC~\cite{refinedc-automating-foundational-sammler-2021}.
On the other hand, dynamic languages such as JavaScript need to reason about both the presence and absence of object properties, in a setting where these properties can be added and deleted throughout program execution. In this case, the object domain is the set of all present properties, with properties not in that set being implicitly treated as absent, as in the work on JaVerT 2.0~\cite{javert-20-compositional-fragososantos-2019}.

We design a map-with-domain combinator, $\pmapdomstm(\type, D, \cmm)$, where $D$ captures the semantics of domains, which can be used as a basis for both use cases.
We first provide the definition of the combinator and then show how it can be instantiated to either CompCert-like static domains or dynamic domains required by JavaScript and other similar languages such as Python, where for the second use case we show that we are required to introduce a specialised version of the combinator, which we call $\dynmapstm(\type, \cmm)$, to capture the intricacies~of~dynamic~languages.

\subsubsection*{Definition} Given a type of addresses $\tau$, a domain $D = (\textsf{ty}, \textsf{in}, \textsf{extend})$ is a triple where:
\begin{itemize}
\item $\textsf{ty}$ is the carrier type of the domain;
\item $\textsf{in} : \type \times D.\textsf{ty}$ is a relation capturing when an address belongs to a domain; and 
\item $\textsf{extend} : \type \to D.\textsf{ty} \to D.\textsf{ty}?$ is a function that extends the domain with an element we know is not in the domain, which may fail (by returning $\None$) if the extension is not allowed. 
\end{itemize}
Now, the combinator $\pmapdomstm(\type, D, \cmm)$ is defined as $\pmapstm(\type, \cmm) \prodstm \exclstm(D.\textsf{ty})$, so that its memories are pairs $(\mu, d)$ of a map $\mu$ and a domain $d \in D.\textsf{ty}?$; the combinator overrides well-formedness, composition, and actions as follows:
\begin{itemize}
\item in addition to the well-formedness of its components, the memory $(\mu, d)$ must meet a map-domain invariant, stating that, when the domain is owned, all keys in the map are $\textsf{in}$ it: $d \neq \None \land k \in \dom(\mu) \implies D.\mathsf{in}(k, d)$;
\item
composition must preserve well-formedness, which requires making sure that domains being composed respect each other's keys; and
\item lifted actions that access keys outside the domain can only do so if the domain can be extended.
\end{itemize}

\subsubsection*{Static domain}

For this presentation, we define a simplified CompCert-like memory model which we refer to as the \emph{block-offset memory model} (BO), originally introduced by \citet{gillian-part-ii-maksimovic-2021}.
This model is currently implemented in Gillian and used in our evaluation (\Cref{sec:eval}).
The memories of the BO model are maps from block identifiers to blocks, where each block is either a pair of an offset-value map and its bound, or freed, i.e., $\cmemss = \locs \fparrow (((\nats \fparrow \vals) \times \nats?) \uplus \{ \freed \})$.

First, we define a domain that captures integer bounds that cannot be extended as follows: $D_< \defeq (\nats, <, \lambda a,d. \None)$. With that in place, we can then capture the BO model (using the lazy map combinator at the top, like the Gillian-JS memory model as discussed previously):
\[
	\pmapconsstm(\locs, \freeablestm(\pmapdomstm(\nats, D_<, \exclstm(\vals))))
\]

For instance, allocating a new block of size $n$ creates a block with $n$ cells at offsets from $0$ to $n-1$ and bound $\Some(n)$. Trying to load from offset $n$ or greater fails, as the offset is not in the domain, and deallocating a block requires ownership of all of the cells \emph{and} of the bound. 
This behaviour is implemented in other C verification tools, such as VeriFast and RefinedC, with the difference that our model also includes erroneous accesses ($\oxerr$ and $\oxm$), which are required for bug-finding.

\subsubsection*{Dynamic domain}

In dynamic languages such as JavaScript, one also must reason about the \emph{absence} of fields within an object, as shown in earlier work~\cite{program-logic-javascript-gardner-2012,javert-javascript-verification-fragososantos-2017,javert-20-compositional-fragososantos-2019}, as fields can be added and removed at runtime. In particular, in order for the reasoning to be sound, one must distinguish between fields that are known to be absent (e.g., if the field was never written to) and fields that are missing (e.g., due to being framed off).
In JavaScript, for example, reading from the former returns \texttt{undefined}, whereas reading from the latter is a logical error that yields a $\oxm$~outcome.

During our work, we have discovered an imprecision in the concrete JavaScript memory model described in~\citet{gillian-part-ii-maksimovic-2021}. 
The issue is that this CMM, unlike the one presented in~\citet{javert-javascript-verification-fragososantos-2017}, represents absent object fields and domains separately, mirroring the SMM. This separation makes it impossible to prove the following, required, semantic equality:
\begin{equation}%
\label{eq:js-dom-undefined}
\models \mathsf{dom}(o, \textit{d}) \lstar p \in \strings \lstar p \notin \textit{d} \iff \models \mathsf{dom}(o, \{ p \} \uplus \textit{d}) \lstar (o, p) \mapsto \freed
\end{equation}
where $\mathsf{dom}(o, \textit{d})$ denotes that object $o$ has domain \textit{d} and $(o, p) \mapsto \freed$ denotes that field $p$ of object $o$ is absent. The fix is straightforward: one must revert, in the concrete case, to the CMM of~\citet{javert-javascript-verification-fragososantos-2017}, where the memory is represented as a partial (but not necessarily finite) mapping from object properties to their values, allowing the domain to be expressed as an infinite union of absent properties.

Guided by this insight, we define
$
\dynmapstm(\type, \cmm) \eqdef \pmapdomstm(\type, D_\in^\type, \freeablestm(\cmm))
$, where
$D_\in^\type = (\mathcal{P}(\type), \in, \lambda s, d.~\Some(\{s\} \uplus d))$
and where the entire CMM is overridden to be based on partial (rather than finite-partial) maps. In particular, the resource $\mathsf{dom}(o, \textit{d})$ is modelled by $(\lambda a.~\text{undefined when } a \in \textit{d} \text { otherwise } \freed)$. We have shown in Rocq that this definition of $\dynmapstm$ does satisfy \Cref{eq:js-dom-undefined}.

We illustrate some of the idiosyncrasies of this memory model in the diagram below.
Firstly, allocating a fresh object yields an empty field map and an empty domain (centre $\leftarrow$ right).
Then, when trying to write to a field $\sym p$ in this empty object, both the domain and the field map are extended to include~$\sym p$ (left $\leftarrow$ centre). If $\sym p$ is later deleted, the field is set to $\freed$.
When such deletion happens, the memory model can exploit the left direction of \Cref{eq:js-dom-undefined} to normalise the memory: fields set to $\freed$ in the map can be removed from both the map and the domain, which can be achieved by a custom action that is similar to the consolidation action of \Cref{sec:consolidation} (left $\rightarrow$ centre).

\vspace{.3cm}
 \begin{minipage}[t]{0.40\textwidth}
 \begin{tikzpicture}[scale=0.6, every node/.style={scale=0.8}]
   \draw (0,0) rectangle (1,3);

   \draw(2,2.5) rectangle (3,3.5);

   \draw[dashed] (2,1) rectangle (8.3,2);

   \draw (2,-0.5) rectangle (3,.5);

   \draw[dashed] (2.2,1.2) rectangle (5.2,1.8);
   \draw (3,1.2) rectangle (4.4,1.8);

   \foreach \y in {1,2} {
     \draw (0,\y cm) -- (1cm,\y cm);
   }

     \node (n3) at (0.5cm, 0.5cm) {$l_3$};
     \node (n2) at (0.5cm, 1.5cm) {$l_2$};
     \node (n1) at (0.5cm, 2.5cm) {$l_1$};

     \node[scale=1.25] (cons) at (6.5cm, 3cm) {$\xrightarrow{\afree([l_2, \sym p]) + \mathsf{consolidate}([])}$};
     \node[scale=1.25] (prod) at (7.5cm, 0cm) {$\xleftarrow{\amutate([l_2, \sym p, 42])}$};

     \node (uthree) at (2.5,3) {...};

     \node (bound) at (6.7,1.5) {\small  $d: \{ \sym p \}$};
     \node (mfalse) at (3.7,1.5) {\small $\sym p : 42$};

     \node (bot) at (2.5,0) {...};

     \draw[-{Stealth[length=2mm]},shorten <=1mm]  (n1.east) -- (2,2.9);
      \draw[-{Stealth[length=2mm]},shorten <=1mm]  (n2.east) -- (2,1.5);
       \draw[-{Stealth[length=2mm]},shorten <=1mm]  (n3.east) -- (2,0);
 \end{tikzpicture}
 \end{minipage}
 \begin{minipage}[t]{0.40\textwidth}
 \begin{tikzpicture}[scale=0.6, every node/.style={scale=0.8}]
   \draw (0,0) rectangle (1,3);

   \draw(2,2.5) rectangle (3,3.5);

   \draw[dashed] (2,1) rectangle (8.3,2);

   \draw (2,-0.5) rectangle (3,.5);

   \draw[dashed] (2.2,1.2) rectangle (5.2,1.8);


   \foreach \y in {1,2} {
     \draw (0,\y cm) -- (1cm,\y cm);
   }

     \node (n3) at (0.5cm, 0.5cm) {$l_3$};
     \node (n2) at (0.5cm, 1.5cm) {$l_2$};
     \node (n1) at (0.5cm, 2.5cm) {$l_1$};

     \node[scale=1.25] (prod) at (7.3cm, 0cm) {$\xleftarrow{\anew([])}$};

     \node (uthree) at (2.5,3) {...};

     \node (bound) at (6.7,1.5) {\small  $d: \emptyset$};

     \node (bot) at (2.5,0) {...};

     \draw[-{Stealth[length=2mm]},shorten <=1mm]  (n1.east) -- (2,2.9);
      \draw[-{Stealth[length=2mm]},shorten <=1mm]  (n2.east) -- (2,1.5);
       \draw[-{Stealth[length=2mm]},shorten <=1mm]  (n3.east) -- (2,0);
 \end{tikzpicture}
 \end{minipage}
 \begin{minipage}[t]{0.20\textwidth}
 \begin{tikzpicture}[scale=0.6, every node/.style={scale=0.8}]
   \draw (0,0) rectangle (1,1);
   \draw[dashed] (0,1) rectangle (1,2);
   \draw (0,2) rectangle (1,3);

   \draw(2,2.5) rectangle (3,3.5);

   \draw (2,-0.5) rectangle (3,.5);

     \node (n3) at (0.5cm, 0.5cm) {$l_3$};
     \node (n1) at (0.5cm, 2.5cm) {$l_1$};

     \node (uthree) at (2.5,3) {...};

     \node (bot) at (2.5,0) {...};

     \draw[-{Stealth[length=2mm]},shorten <=1mm]  (n1.east) -- (2,2.9);
       \draw[-{Stealth[length=2mm]},shorten <=1mm]  (n3.east) -- (2,0);
 \end{tikzpicture}
 \end{minipage}%
 
\section{Engine Combinators}%
\label{sec:zoo-engine}

We give two examples of how combinators can be used beyond modelling memory, capturing two concepts commonly hard-coded into the core of a CSE engine: user-defined predicates (which are supported by all CSE verifiers, including VeriFast, Viper, Gillian, and CN); and bi-abductive analysis (supported by Gillian and Infer-Pulse). In addition to facilitating modularity, this allows one to easily implement variants of these traditional components that may be required by advanced analysis applications, as illustrated in our discussion of the Gillian-Rust memory model in \Cref{sec:eval}.

\subsection{User-defined Predicates as a Combinator}%
\label{sec:zoo-predicates}
 
\newcommand{\lstvs}{\lvar{vs}}

All CSE verifiers support user-defined predicates using a symbolic execution technique called \emph{iso-recursive predicates}~\cite{formal-semantics-isorecursive-summers-2013}, where the symbolic semantics of predicates is handled differently between the predicate instance (when it is \emph{folded}) and its body (when it is \emph{unfolded}).
We introduce the \emph{predicate combinator}, $\predstm(\predset, \cmm)$, which factors out this way of handling predicates in $\predset$ into a reusable self-contained component that can be applied to any memory model $\cmm$. Its implementation in Gillian illustrates how intricate automation mechanisms (such as auto-folding and unfolding), previously orchestrated by the core engine, can be delegated to a combinator, showcasing the flexibility of our approach.
There is no counterpart to $\predstm$ in any previous work on constructing PCMs for program logics using combinators, as the combinator is only relevant for CSE tools.

The complexity of this combinator arises from the fact that predicate resources can occur in predicate bodies. In more detail, to give the type of the parameter $\predset$, we first need to define the combinator resources: $\resources \defeq \{ \rsub(\resource) \mid \resource \in \resources^\cmm \} \cup \{ \rpred(\pred) \mid \pred \in \strings \}$, containing resources lifted from~$\cmm$ and the predicate names themselves. Now, the parameter $\predset : \strings \parrow \vec{\LVar} \times \vec{\LVar} \times \Assert^\resources$ describes, for each predicate name $\pred$, a triple $\predset(\pred) = (\vec{\lvar{x}_i},\vec{\lvar{x}_o}, A)$ corresponding to the in- and out-parameters of the predicate, and its body $A$, an assertion from the full assertion language instantiated with resources $\resources$. We require that predicate bodies be of the form $A = \bigvee_{i \in I} A_i$.

\begin{example}\label{ex:linkedlist} Given the linear heap $\pmapstm(\nats, \exclstm(\val))$, the standard linked-list predicate of SL is added using $\predstm(\predset_{\text{list}}, \pmapstm(\nats, \exclstm(\val)))$ where $\predset_{\text{list}} = \{ \mathit{list} \mapsto ([\lvar{x}], [\lvar{vs}], A_\text{list}) \}$ and
\[
\begin{array}{@{}r@{~}c@{~}l@{}}
A_\text{list} & \defeq & (x\doteq\nil \lstar \lstvs \doteq \emplist) \lor (\exsts{v, x', \lstvs'} x \mapsto v, x' \lstar \rpred(\mathit{list})([x']; [\lstvs']) \lstar \lstvs \doteq v \cons \lstvs')
\end{array}
\]
where the points-to assertion $x \mapsto v, x'$ (meaning $x \mapsto v \lstar x + 1 \mapsto x'$) is lifted from the~linear~heap.
\end{example}

The set of actions of the combinator is $\actions \defeq \{ \asub(\action) \mid \action \in \actions^\cmm \} \cup \{ \afold, \aunfold \}$, that is, the lifted actions $\asub(\action)$, which execute an action $\action$ of $\cmm$, together with two new actions, $\afold$ and $\aunfold$, which move predicates between their folded and unfolded state, as we describe below.

\subsubsection*{Concrete combinator} As predicates are symbolic-only constructs, the concrete predicate combinator $\predstm(\predset, \cmm)$ is mostly trivial: it leaves the set of concrete memories unchanged, $\cmemss^{\predstm(\predset, \cmm)} \defeq \cmemss^\cmm$, where concrete $\aunfold$ and $\afold$ are no-ops.

The only interesting definition in the concrete combinator is the satisfaction relation for resources $\resmodels$: as predicate bodies can reference predicates (including the predicate itself), it must be defined by mutual recursion with the satisfaction relation for the full assertion language. In more detail, let $\asrtmodels^{\resources}$ denote the satisfaction relation for the full assertion language $\Assert^\resources$ instantiated with $\resmodels$. Then, $\resmodels$ itself is defined as follows:
\begin{mathpar}\small
\inferrule
 {\cmem \resmodels^\cmm \resource(\vin; \vout)}
 {\cmem \resmodels \rsub(\resource)(\vin; \vout)}
\and
\inferrule
 {\predset(\pred) = (\xin, \xout, A) \and \{ \xin \mapsto \vin; \xout \mapsto \vout \}, (\emptyset, \cmem) \asrtmodels^{\resources} A}
 {\cmem \resmodels \rpred(\pred)(\vin; \vout)}
\end{mathpar}

\subsubsection*{Symbolic combinator} The symbolic predicate combinator $\spredstm(\predset, \smm)$ has symbolic memories $\smemss \defeq \smemss^{\smm} \times \predlists$ that are pairs of a symbolic memory of the input memory model ($\smemss^{\smm}$) and a \emph{multiset of folded predicates} $\sym{P} \in \predlists$, with elements of the form $\pred(\sexpin; \sexpout)$.

The semantics of lifted actions $\asub(\action)$ is straightforward except for one subtlety for $\oxm$ outcomes:
\begin{mathpar}\small
\inferrule
 {\smem.\action(\sexps, \spc) \baction_m^{\smm} \outcome : (\sexps', \smem', \spc') \and
  \outcome' = \oxabort \text{ if } (\outcome = \oxm \text{ and } \sym{P} \neq \emptyset) \text{ else } \outcome}
 {(\smem, \sym{P}).\asub(\action)(\sexps, \spc) \baction_m^{\spredstm(\predset, \smm)} \outcome' : (\sexps', (\smem', \sym{P}), \spc')}
\end{mathpar}
which tells us that $\oxm$ outcomes must be reported as $\oxabort$ outcomes in the presence of predicates, as not doing so would break both OX and UX soundness, given that the missing resource may be hidden inside one of the predicates.

Intuitively, the $\afold$  action \emph{adds} a folded predicate to the predicate multiset and \emph{consumes} the body of the predicate from the memory, whereas $\aunfold$ \emph{produces} its body and \emph{removes} the folded predicate from the multiset. The most interesting part of the definition of the two actions arises from the recursive nature of the predicates. Because predicate bodies are assertions which can contain the predicate itself and other predicates, the definition of the actions must be aware of the consume and produce operations for the full assertion language instantiated with the predicate resources. To illustrate, the successful rule for the unfold action is as follows, where $\produce_{\mathcal{A}}^{\resources}$ denotes the assertion-level produce operation instantiated with $\resproduce$:
\begin{mathpar}\small
\inferrule
 {\predset(\pred) = (\xin, \xout, \bigvee_{i \in I} A_i) \\ \text{angelically pick } \pred(\sexpin'; \sexpout') \in \sym{P} \text{ s.t. } \spc \models \sexpin' = \sexpin \and \sym{P}' = \sym{P} \setminus \{ \pred(\sexpin'; \sexpout') \} \\\\ (\smem, \sym{P}').\produce_{\mathcal{A}}^{\resources}(A_i, \{ \xin \mapsto \sexpin'; \xout \mapsto \sexpout' \}, \spc \land \sexpin' \in \vals \land \sexpout' \in \vals) \baction ((\smem', \sym{P}''), \spc')}
 {(\smem, \sym{P}).\aunfold(\pred :: \sexpin, \spc) \baction_m^{\spredstm(\predset, \smm)} \oxok : ([], (\smem', \sym{P}''), \spc')}
\end{mathpar}

\smallskip\noindent
\textit{Example~\ref*{ex:linkedlist} (continued).}~ Using the same memory model, $\spredstm(\predset_{\text{list}}, \spmapstm(\nats, \sexclstm(\vals)))$, take the memory $\smem_0 = (\emptyset, \{ \mathit{list}([\sym{x}]; [\sym{vs}]) \})$. In this initial state, trying to access the cell at address $\sym{x}$ would abort, because the heap (first component) is empty. In order to access the cell, we first need to \emph{unfold} the predicate, executing the action $\aunfold([\mathit{list}, \sym{x}])$. The action receives the name and in-parameters of the predicate as arguments, identifies and removes the corresponding element in the multiset of folded predicates, and then branches to produce both disjuncts of its body into the symbolic memory, resulting in the following, where $\sym{y}_{\{1, 2, 3\}}$ are fresh symbolic variables:
\[\small
\begin{array}{ll}
\smem_1 = (\emptyset, \emptyset) & \spc_1 = (\sym{x} = \nil) \land (\sym{vs} = \emplist) \\
\smem_2 = (\{ \sym{x} \mapsto \sym{y}_1; \sym{x} + 1 \mapsto \sym{y}_2 \}, \{ \mathit{list}([\sym{y}_2]; [\sym{y}_3]) \}) & \spc_2 = (\sym{vs} = \sym{y}_1 \cons \sym{y}_3)
\end{array}
\]

\subsubsection*{Soundness} We have proven the symbolic combinator both OX- and UX-sound, with the exception of the $\afold$ action, which is only OX-sound. The soundness proofs and related definitions are more involved than for other combinators because of the above-discussed recursive nature of predicates.

\subsubsection*{Implementation} Our implementation of the $\predstm$ combinator in Gillian was a substantial refactoring; despite its scale, we managed to recover all existing case studies, including the most fragile proofs, which largely relied on the nuances of the automation mechanisms for predicates.

Beyond the presented formalisation, the Gillian predicate combinator has a number of additional features that have proven to be essential for efficient real-world verification, such as the $\mathsf{rec\_unfold}$ action that unfolds a predicate recursively until a base case is reached or fuel is exhausted. We do not formalise these actions as they bring no additional insight: for example, the  soundness of $\mathsf{rec\_unfold}$ is a direct consequence of the soundness of $\aunfold$.

Similarly, when an action or consumption fails, the Gillian engine may attempt recovery by folding or unfolding a predicate, the choice of which is guided by empirically established heuristics. In our $\predstm$ combinator, this is implemented by wrapping action evaluation and resource consumption in these recovery mechanisms instead of hard-coding them in the engine, again showing just how flexible the combinator approach is.
We also expose an additional $\mathsf{recover}$ action, similar to consolidation, which can be called anywhere in the engine to insert such recovery attempts. The soundness of the recovery mechanism is a direct consequence of the soundness of $\afold$ and $\aunfold$.

 \subsection{Bi-abduction as a Combinator}
 \label{sec:zoo-biabduction}

Bi-abduction~\cite{compositional-shape-analysis-calcagno-2009,local-reasoning-presence-raad-2020} is a program analysis technique for incrementally discovering the resources required to execute a program fragment. It is used in CSE engines such as Infer-Pulse~\cite{finding-real-bugs-le-2022} and Gillian~\cite{compositional-symbolic-execution-loow-2024} to infer missing pre-conditions and enable compositional bug-finding. 

Like the predicate combinator, the bi-abduction combinator operates purely at the symbolic level. 
Its symbolic memories are pairs $(\smem, \antif)$, comprising an underlying memory $\smem \in \smemss^{\smm}$ and an \emph{anti-frame} $\antif$, which is a list of resource descriptions of the form $(\resource, \sexpin, \sexpout)$ where $\resource \in \resources^{\smm}$. This anti-frame represents resources that were missing from the initial symbolic memory but are required for the execution of the analysed code, and it makes the bi-abduction combinator neither OX- nor UX-sound in the standard sense. 
Instead, intuitively, the combinator guarantees that bi-abduction infers valid ISL specifications: that is, if executing a command $\cmd$ starting from the initial memory $(\emptysmem, [])$ results in a final memory $(\smem, \antif)$ with outcome $\outcome \in \{ \oxok, \oxerr \}$, then the ISL triple $\isltriplex{ \mathtt{toAsrt}(\antif) }{ \cmd }{\outcome}{ \mathtt{toAsrt}(\smem) }$ holds, where $\mathtt{toAsrt}$ converts a symbolic memory or anti-frame to an ISL assertion in $\Assert^{\resources}$. This is captured formally (and proven in Rocq) as follows:
\[
\begin{array}{l}
\smm \soundux \cmm \land 
\cseenginetrans{\sbistm(\smm)}{\fsctx}{\sst}{\scmd}{\outcome}{\sst'} \land 
\outcome \in \{ \oxok, \oxerr \} \land
\subst, \st' \models \sst' \land
\sst.\mathtt{mem} = (\_, []) \land
\sst' = (\_, (\_, \antif), \spc') \\
\quad \implies
\exists \sto, \cmem, \cmem_\text{fix}.~
\subst, (\sto, \cmem) \models \sst[\mathsf{pc} := \spc'] \land \subst, \cmem_\text{fix} \asrtmodels \mathtt{toAsrt}(\antif) \land (\sto, \cmem \memcomp \cmem_\text{fix}), \scmd \baction_{\fictx}^{\bistm(\cmm)} \outcome : \st'
\end{array}
\]

The implementation of the combinator is in the style of \emph{fixes-from-errors} bi-abduction~\cite{javert-20-compositional-fragososantos-2019,compositional-symbolic-execution-loow-2024,compositional-symbolic-execution-loow-2025}. At the level of memory actions, the main rule is as follows:
 \[\small
 \inferrule{
 \smem.\action(\vec{\sexp}, \spc) \baction^{\smm} \outcome_m : (\vec{\sexp}_m, \smem_m, \spc_m) \and
 \outcome_m \in \{ \oxm, \oxabort \} \and
 \fixfn(\vec{\sexp}_m) = \resource(\sexpin; \sexpout) \\\\
 \smem.\resproduce^{\smm}(r, \sexpin, \sexpout, \spc_m) \baction^{\smm} (\smem', \spc') \\
 (\smem', \resource(\sexpin; \sexpout) \cons \antif).\action(\vec{\sexp}, \spc') \baction^{\sbistm(\smm)} \outcome : (\vec{\sexp}_f, (\smem_f, \antif_f), \spc_f)
 }{
 (\smem, \antif).\action(\vec{\sexp}, \spc) \baction^{\sbistm(\smm)} \outcome : (\vec{\sexp}_f, (\smem_f, \antif_f) , \spc_f)
 }\]
 
 Let us walk through this rule. When an action $\action$ of $\smm$ yields a $\oxm$ or $\oxabort$ outcome, the error value is passed to a $\fixfn$ operation, which is a combinator-custom operator that must be provided by the input memory model. The operation $\fixfn$ proposes a resource that may correct the error. This resource is then produced into the initial state $\smem$ to obtain a ``fixed'' memory $\smem'$, and the fix is added to the anti-frame, indicating that it was a missing pre-condition. Finally, the action is replayed in $\smem'$ \emph{at the combinator level} (note the $\sbistm(\smm)$ in the last premise), so that, should the replay miss or abort again, further resources are abduced in turn until the action succeeds, yielding the final state $\smem_f$ and accumulated anti-frame $\antif_f$.

There are several noteworthy remarks to make. First, the rule above is simplified for readability in two ways: the $\fixfn$ operation should in practice not just be allowed to return a single resource but multiple resources; and the rule needs a fuel parameter to limit the number of attempted fixes, guaranteeing that it does not loop forever. Second, it is not necessary to impose a soundness requirement on the $\fixfn$ operation for the bi-abduction to be sound: if the $\fixfn$ operation proposes an invalid fix, execution will simply fail to continue (that is, no specification will be inferred), or create an unusable specification (equivalent to $\isltriplex{ \mathit{False} }{ \cmd }{\outcome}{ \mathit{False} }$), which is not desirable but is also not unsound. Finally, constructing bi-abduction using a combinator, which replays only the latest memory operation in case of $\oxm$ or $\oxabort$ (as in JaVerT 2.0), is more efficient than the approach of \citet{compositional-symbolic-execution-loow-2024}, which replays execution of the entire command.

%% file: sections/eval.tex

\section{Evaluation}%
\label{sec:eval}

We re-implement existing demonstrator and real-world monolithic memory models using CSE à la carte and evaluate our Rocq mechanisation and our OCaml implementation for Gillian.
For Rocq, we prove, for a selection of demonstrator memory models, that the monolithic and combinator-based models are behaviourally equivalent; we cannot evaluate the real-world models of Gillian in this way, as their monolithic versions have not been mechanised in previous work and there is therefore nothing to compare against.
For Gillian, we give the full details of combinator-based models for JavaScript and C, sketch the one for Rust, and then compare the behaviour, size, and performance of all implemented monolithic vs. combinator-based models.
We report that the Gillian combinator-based models do not exhibit any unexpected behaviour across all previously published verification and testing case studies, are smaller in size than their monolithic counterparts, and introduce only a negligible performance overhead.

\subsection{Rocq Models: Theoretical Evaluation}%
\label{sec:eval-theory}

Using our Rocq combinators, we have re-implemented a representative subset of the monolithic memory models (MMs) from \citet{compositional-symbolic-execution-loow-2025}'s Rocq mechanisation. In particular, we re-implement:
\begin{itemize}
\item an OX-only linear MM with exclusive ownership as $\pmapstm(\nats, \freeableoxstm(\exclstm(\val)))$;
\item an exact linear MM with fractional ownership as $\pmapconsstm(\nats, \freeablestm(\fracstm(\val)))$; and
\item an exact block-offset MM as $\pmapconsstm(\nats, \freeablestm(\pmapdomstm(\nats, D_<, \exclstm(\vals))))$.
\end{itemize}

We validate our combinator-based memory models by proving in Rocq that they are concretely and symbolically behaviourally equivalent to the existing monolithic models. Specifically, we show that all three combinator-based CMMs are behaviourally equivalent to the existing models for all $\oxok$ outcomes precisely, and for all $\oxerr$ and $\oxm$ outcomes \emph{up to the order in which errors are checked}. Further, we establish that the SMM of our combinator-based OX-only linear memory model is behaviourally equivalent to that of the existing monolithic memory model, modulo certain instances of differences in the order of branching on errors, which is subtly different from the order of error checking. We summarise our results below, with further details in~\appref{app:app-eval}{B} and Park's PhD thesis (in preparation).

\subsubsection*{Concrete equivalence} It is important that our combinator-based CMMs behave as expected as they constitute the trusted computing base for our  soundness proofs. Our setup for comparison is as follows: we say a CMM $\cmm_1$ is simulated by another CMM $\cmm_2$, denoted by $\cmm_1 \sqsubseteq \cmm_2$, iff, informally, every memory action step in $\cmm_1$ has a corresponding step in $\cmm_2$. Formally, we say that $\cmm_1 \sqsubseteq \cmm_2$ iff there exist relations $\simeq_\cmemss$, $\simeq_\actions$, $\simeq_{\emph{ins}}$, and $\simeq_{\emph{outs}}$ (formally capturing ``corresponding'') such that:
\[
\begin{array}{l}
\hp_1.\alpha_1(\vinone) \baction \outcome: (\voutone, \hp_1') \land \hp_1 \simeq_\cmemss \hp_2\ \land \cwf_1(\hp_1) \implies \\
\quad\exists \hp_2', \vouttwo, \alpha_2, \vintwo. \\
\quad\quad\action_1 \simeq_\actions \action_2 \land
\vinone \simeq_{\emph{ins}} \vintwo \land \hp_1' \simeq_\cmemss \hp_2' \land \voutone \simeq_{\emph{outs}} \vouttwo \land  \hp_2.\alpha_2(\vintwo) \baction \outcome: (\vouttwo, \hp_2')
\end{array}
\]
where these correspondences are instantiated as follows for the OX-only linear memory model:
\begin{itemize}
\item $\hp_1 \simeq_\cmemss \hp_2  \mathit{~iff~} \dom(\mu_1) = \dom(\mu_2) \land (\forall a.\ a \in \dom(\mu_1) \implies \mu_1(a) = \Some (\mu_2(a)))$;\footnote{Note that we do not have a strict equality between the two memories. This is because the memory structure of the combinator model is $\nats \fparrow \val?$, whereas the monolithic variant $\cmmlin$ has the memory structure $\nats \fparrow \val$. Due to the mismatch in the codomain of the memories, a $\Some$ wrapper is required.}
\item actions are related straightforwardly: for example, $\asub(\asub(\amutate))$ relates to $\amutate$, etc.;
\item $\lst v_1 \simeq_{\emph{ins}} \lst v_2 \mathit{~iff~} \lst v_1 = \lst v_2$; and
\item $\lst v_1 \simeq_{\emph{outs}} \lst v_2$ also takes into account the mapping of errors from one memory model to the other.
\end{itemize}
We say that two CMMs, $\cmm_1$ and $\cmm_2$, are behaviourally equivalent iff $\cmm_1 \sqsubseteq \cmm_2$ and $\cmm_2 \sqsubseteq \cmm_1$. As stated above, we have shown our combinator-based models to be behaviourally equivalent (using tight relations $\simeq_\cmemss, \simeq_\actions, \dots$) to the existing models up to differences in the order of error checking, e.g., one model checking the length of inputs before checking the types of the inputs, which arise from the fact that error checking must be done combinator-by-combinator in the combinator stacks rather than once as in the monolithic memory models. An in-depth explanation for the OX-only linear memory model is given in~\appref{app:app-eval}{B}.

\subsubsection*{Symbolic equivalence} The behavioural equivalence between two SMMs, also defined by a two-way simulation relation (see \appref{app:app-eval}{B} for the formal definition), is a stronger and more intricate relation than the one for CMMs. The relationship is stronger in the sense that two SMMs can be sound w.r.t. the same CMM but not be behaviourally equivalent to each other because they might differ in how they branch and how much they over/under-approximate. The relationship is more intricate because there are more moving parts than in concrete execution: for symbolic actions, we must keep track not only of memories but also of path conditions, cut unsatisfiable paths before comparing outcomes, and compare the consume and produce operations. In comparing the two OX-only linear memory models, the only behavioural difference that we found is that the number of branches differs in some situations as a result of the two models branching over errors in a different order (for example, checking the length of input before checking if a memory access is out of bounds and vice versa), which arises because branching, like error checking, is done combinator-by-combinator.

\subsection{Gillian Models: Empirical Evaluation}%
\label{sec:eval-emp}

We confirm that our approach scales to real-world analysis tools by \begin{enumerate*}
\item implementing in Gillian all components presented in this paper;
\item developing utility combinators to adapt combinator-based models to existing monolithic interfaces (e.g., for renaming actions);
\item using these components to re-implement a number of existing Gillian memory models, most notably those of Gillian-JS and Gillian-C~\cite{gillian-part-multilanguage-fragososantos-2020,gillian-part-ii-maksimovic-2021};
\item providing a sketch of a combinator-based construction for the Gillian-Rust memory model~\cite{hybrid-approach-semiautomated-ayoun-2025}, not implemented due to manpower constraints only; and
\item successfully replaying all relevant verification and testing case studies of Gillian using the combinator-based~models.
\end{enumerate*}

\subsubsection*{Real-world memory-model construction in Gillian}%
We discuss what it means to construct the memory models of  JavaScript, C, and Rust in Gillian using our combinator-based approach.

\paragraph{Gillian-JS}

In theory, the combinator-based Gillian-JS memory model should be as described in \Cref{sec:overview} except that the top-level map combinator should be lazy (a detail not introduced in the overview):
\[
\pmapconsstm(\locs, \dynmapstm(\strs, \exclstm(\vals)) \prodstm {\agstm(\locs)})
\]
mapping object locations to objects and their metadata, where each object is a dynamic map from string properties to values.
The Gillian implementation closely follows this structure, and adds some general-purpose utility combinators and patches (shaded in \shade{\text{grey}} below) to re-establish with total faithfulness the interface and behaviour of the original monolithic model:
\[
\shade{\stmfont{Filter}_{\jslang}}(
    \shade{\stmfont{Mapper}_{\jslang}}
        (\shade{\pmapstm_{\jslang}}(\locs,
                \shade{\stmfont{InToOut}_{\jslang}}(
                \dynmapstm(\strs, \exclstm(\vals)))
                \shade{\prodstm_{\jslang}} {\agstm(\locs)})
    )
)
\]
For the utility combinators: $\stmfont{Filter}_{\jslang}$ removes unused actions/resources to match the monolithic interface;
$\stmfont{Mapper}_{\jslang}$ renames actions/resources to match the monolithic model (e.g., $\inl(\alookup)$ becomes $\agetprop$);
$\pmapstm_\jslang$ patches $\pmapstm$ to allocate an object and its metadata in a single action, and implements custom logic for a recovery tactic (explained shortly);
$\stmfont{InToOut}_{\jslang}$ modifies the JS object resource signature to match the original model (changing out-parameters to in-parameters); and
$\prodstm_{\jslang}$ patches allocation initialisation to optionally receive a metadata argument.
The glue code for these changes is only 302 lines of OCaml, roughly a third of the size of the original model, while preserving behavioural equivalence, as confirmed by the Gillian-JS case studies reported below.

Importantly, as the core model construction is straightforward, developers are left free to focus on JS-specific reasoning rather than boilerplate. For example, one crucial automation of Gillian-JS, a recovery tactic that finds an object location from its metadata on failure, is implemented as a small, non-soundness-critical patch of only 30 lines.

We conclude with a note on UX reasoning. While the Gillian-JS model was originally designed for OX reasoning, constructing the model using our combinators shows that the model is also UX-sound, with one minor wrinkle: because of the $\agstm(\locs)$ component in the codomain of the $\pmapconsstm$ combinator, new objects cannot be allocated because no element of $\agstm(\locs)$ is exclusive. This wrinkle is easily ironed out by replacing $\agstm(\locs)$ with a mode-specific read-only memory model, $\rostm_m$, which gives shared read-only ownership under $\mOX$ and exclusive read-only ownership under $\mUX$, and can safely be used as a partial-map codomain. In more detail, $\rostm_m$ is defined as ``$\text{if } m = \mOX \text{ then } \agstm \text{ else } (\exclstm \text{ without } \amutate)$'', where $\amutate$ is removed to prevent updates.

\paragraph{Gillian-C}%

\newcommand{\typeC}{\type_\text{C}}

In Gillian-C~\cite{gillian-part-multilanguage-fragososantos-2020,gillian-part-ii-maksimovic-2021,gillian-foundations-implementation-ayoun-2024}, memories are pairs of a heap and a global environment, where heaps map block identifiers to memory blocks represented using a complex tree-based structure described by \citet{gillian-foundations-implementation-ayoun-2024}; and global environments map function pointers to their definitions, enabling dynamic function calls. Using combinators and an additional custom MMA, $\treeblockstm_{\clang}$, we can express the existing Gillian-C memory model as (eliding utility combinators):
\[
\pmapconsstm(\locs, \freeablestm(\treeblockstm_{\clang})) \prodstm \pmapconsstm(\locs, {\rostm_m(\val)})
\]
We re-implemented the Gillian-C memory model following this structure, obtaining a model whose model-specific code is 17\% smaller than the original monolithic code, factoring out soundness- and performance-critical code while preserving full behavioural equivalence.
A large portion of the code still corresponds to $\treeblockstm_{\clang}$, which provides strong automation and good performance for reasoning about C, but is complex and tightly tied to the C memory semantics. We will explore the decomposition of $\treeblockstm_{\clang}$   using combinators and mechanise soundness in future.

\paragraph{Gillian-Rust}
\citet{hybrid-approach-semiautomated-ayoun-2025} introduced Gillian-Rust, an instantiation of Gillian for reasoning about type safety and functional correctness of unsafe Rust code. Its memory model uses advanced ghost state constructs to encode a subset of the lifetime logic of RustBelt~\cite{rustbelt-securing-foundations-jung-2017} and the parametric prophecies of RustHornBelt~\cite{rusthornbelt-semantic-foundation-matsushita-2022}. One such construct, namely guarded predicates, $\stmfont{GdPred}$, required the authors to modify the way in which Gillian works with user-defined predicates, partly defeating Gillian's parametric design. Here, we present a sketch of a combinator-based re-implementation that would not require any modification of Gillian, leaving the actual implementation for future~work.

\begin{wrapfigure}{r}{0.48\textwidth}
\vspace{-0.6cm}
\hspace*{-0.25cm}
\begin{minipage}[t]{0.48\textwidth}
\[\small
\begin{array}{l@{~}c@{~}l}
\cmm_{\rustlang} & \defeq & \stmfont{GdPred}(\rustheaps \prodstm \lftstm \prodstm \obsstm \prodstm \pcystm) \\
\rustheaps & \defeq & \pmapconsstm(\locs, \freeablestm(\treeblockstm_\rustlang)) \\
\lftstm & \defeq & \pmapstm(\lfts, \fracstm(\{ \top \}) \sumstm \agstm(\{ \dagger \})) \\
\obsstm & \defeq & \text{custom, but small} \\
\pcystm & \defeq & \pmapstm(\pcys, \agstm(\vals) \prodstm \exclstm(\{ O \}) \prodstm \exclstm(\{ C \}))
\end{array}
\]
\end{minipage}
\vspace{-1.5em}
\end{wrapfigure}

The definition to the right gives the Gillian-Rust memory model $\cmm_{\rustlang}$ constructed using combinators. Heaps $\rustheaps$, lifetime contexts $\lftstm$, observation contexts $\obsstm$, and prophecy contexts $\pcystm$ are composed using our product combinator and used as input to the guarded predicate combinator $\stmfont{GdPred}$, all of which we briefly describe now.
First, a Gillian-Rust heap is similar to a Gillian-C heap, except that $\treeblockstm_\rustlang$ is adapted to reasoning about Rust structures where the order of fields is not guaranteed.
Second, lifetime contexts $\lftstm$, directly adapted from RustBelt, map lifetime identifiers $\lfts$ to either a fractional permission corresponding to an ``alive'' lifetime token $\alivetkn{\lft}_q$, or an agreement for duplicable dead lifetime tokens $\deadtkn{\lft}$.
Third, the observation context $\obsstm$ acts as a second layer of path condition and is composed of a single boolean symbolic expression, where composition is defined as conjunction, forming a commutative monoid; its implementation is straightforward.
Next, the prophecy context $\pcystm$ maps prophecy identifiers $\pcys$ to their values, as well as two flags indicating ownership of resources called ``observer'' ($O$) and ``controller'' ($C$)~\cite{rusthornbelt-semantic-foundation-matsushita-2022}.
Finally, the whole memory model is wrapped in the guarded predicate combinator $\stmfont{GdPred}$, which behaves like (and replaces) the more standard predicate combinator $\predstm$, and is used to model RustBelt's full borrows (we refer the reader to \citet{hybrid-approach-semiautomated-ayoun-2025} should they want to understand how this is achieved).

\subsubsection*{Evaluation: behavioural equivalence}
We demonstrate that our re-implementations of existing models using combinators are behaviourally equivalent to their original monolithic counterparts by replaying all relevant Gillian case studies: verification exercises designed for the BO memory model used in teaching at SSFT'25~\cite{ssft} and publicly available in the Gillian repository; a small concurrency verification case study for the fractional BO memory model used in previous work~\cite{Bucur23}; a symbolic test suite~\cite{gillian-part-multilanguage-fragososantos-2020} and bi-abductive specification inference~\cite{compositional-symbolic-execution-loow-2024} for the real-world Collections-C data-structure library~\cite{srdja-collectionsc-panic-2024}; full functional verification of industry-grade message header deserialisation code from the AWS Encryption SDK~\cite{aws-encryption-sdk} in C and JavaScript~\cite{gillian-part-ii-maksimovic-2021}; and a symbolic test suite~\cite{gillian-part-multilanguage-fragososantos-2020} for the real-world Buckets.js data-structure library~\cite{BucketsJS2025}.
Note that the specification-inference case study compares combinator-based bi-abduction applied to the combinator-based Gillian-C model against Gillian's built-in bi-abduction applied to the original monolithic Gillian-C model.
Importantly, all experiments, including those with monolithic models, were run with the predicate combinator in place: the comprehensive change from core-engine to combinator predicates involved numerous small changes to memory-model implementations that impacted performance (some positively, some negatively), so comparing against times with core-engine predicates would be misleading, even though they~differ~negligibly. 

\begin{table}[!t]
\caption{Summary of case studies previously used to evaluate Gillian, showing the number of execution branches explored, their size in lines of code, and the runtime of the combinator-based MM (CO) compared to the original monolithic MM (MO).
All experiments are performed in the artefact's Docker image (arm64, 8\,GB RAM) on a MacBook M4 Max with 128\,GB RAM; times are the mean of 5 runs after one warm-up run.}
\label{tab:runtime}
\footnotesize
\vspace*{-0.15cm}
\begin{tabular}{@{} l l l r r r r @{}}
\toprule
Case study & Model & Analysis & Branches (CO/MO) & kLoC & Time CO (s) & Time MO (s) \\
\midrule
SSFT'25~\cite{ssft} & Block-offset & Verification & 108 & 2  & 1.08 & 1.08 \\
Small Concur.~\cite{Bucur23} & Frac. block-offset & Verification & 40  & 0.5 & 0.34 & 0.34 \\
Collections-C~\cite{gillian-part-multilanguage-fragososantos-2020} & C & Symb. testing & 463 & 12  & 10.99 & 11.05 \\
Collections-C~\cite{compositional-symbolic-execution-loow-2024} & C & Spec. inference & 3779/3600 & 6  & 8.3 & 10.4 \\
AWS Enc.~\cite{gillian-part-ii-maksimovic-2021} & C & Verification & 378 & 4   & 113 & 112 \\
Buckets.js~\cite{gillian-part-multilanguage-fragososantos-2020} & JS & Symb. testing & 509/508 & 26  & 20.8 & 19.8 \\
AWS Enc.~\cite{gillian-part-ii-maksimovic-2021} & JS & Verification & 111 & 2 & 9.27 & 9.20 \\
\bottomrule
\end{tabular}
\vspace{-0.3cm}
\end{table}

\Cref{tab:runtime} summarises the case studies, the number of execution branches explored, their size in lines of code, and the runtime of the combinator-based memory models compared to the original monolithic memory models.
For all case studies except two, we observed full parity between the two approaches.
First, one additional branch is explored in one test by the JavaScript combinator-based model in Buckets.js due to an unsoundness in the monolithic model we found during evaluation.
Second, in the bi-abductive case study, the engines generate a slightly different but comparable number of specs.
This is due to the fact that the original bi-abduction implementation in Gillian re-implements parts of the logic of the core engine itself, including function calls, using a slightly different strategy for branch exploration.

The times also indicate that combinator-based memory models are as efficient as monolithic ones: the two run within 1\% of each other on all case studies except two.
On Buckets.js, CO is 5\% slower; profiling shows that the extra time is spent executing memory actions, which go through more layers of indirection with combinators than with a monolithic model. We believe this is an artefact of the current implementation of memory actions in Gillian, and that a more efficient implementation would remove this overhead.
On specification inference, CO is 20\% faster, but, as explained above, this case study also compares two different bi-abduction engines.

\begin{figure}[!t]
\centering
\includegraphics[width=0.6\textwidth]{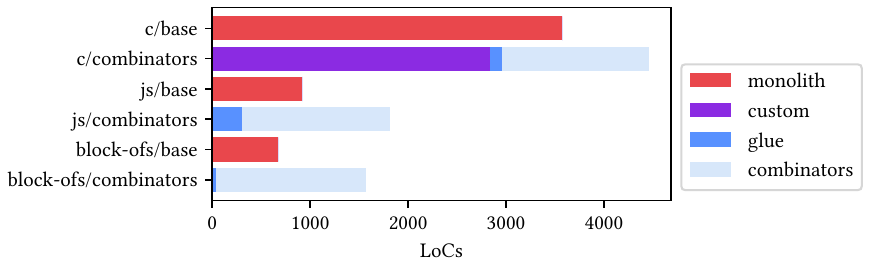}
\vspace*{-0.2cm}
\caption{Lines of code of combinator-based and monolithic memory models, for C, JavaScript, and block-offset.}
\label{fig:loc}
\vspace*{-0.4cm}
\end{figure}

\subsubsection*{Evaluation: memory-model size}%
We measure the number of lines of code of the combinator-based memory models, comparing them to their original monolithic counterparts, demonstrating that our approach systematically leads to smaller memory models.

Both memory models implement the same interface, and we measure that implementation; lines of code exclude whitespace and comments.
These measurements are given in \Cref{fig:loc}, splitting the lines of code of the combinator-based models into: \begin{enumerate*} \item code that implements a custom part of the memory model (only $\treeblockstm_\clang$), \item glue code that instantiates the combinators to build the memory model and possibly adapts behaviour, and \item code that is part of the combinator library. \end{enumerate*}
A significant reduction in code size can be observed for the Gillian-JS and block-offset models, shaving off 67\% and 95\% of the code, respectively.
For Gillian-C, the reduction is more modest but still satisfying at 17\%, mostly because of the $\treeblockstm_\clang$ component.
In addition, our novel version of the $\bistm$ model is composed of 108 LoC performing fairly simple reasoning, and replaces the 423 LoC of bi-abductive reasoning performed as part of the Gillian core, overriding the complex logic behind compositional function~calls.

On a more qualitative note, we observed that writing models using combinators leads to significantly cleaner code, avoiding copy-pasted boilerplate for each action, consume, and produce.

%% file: sections/limitations.tex
\section{Limitations and Threats to Validity}%
\label{sec:limitations}

Our work inherits the limitations fundamental to all CSE tools, such as the reliance on SMT solvers; beyond these, we identify the following limitations and threats to validity.

\subsubsection*{Applicability beyond the CSE theory by Lööw et al.\ and Gillian (\Cref{sec:intro} and \Cref{sec:overview})}
We have shown that our approach works with one specific CSE theory and one specific CSE tool.
The main requirement on a theory or tool for our combinator approach to apply is that it be parametric on the memory model.
The authors of VeriFast have confirmed in private communication that implementing our approach in their tool would be straightforward.
For Viper, our approach would need to be adapted to implicit dynamic frames~\cite{implicit-dynamic-frames-smans-2012}, which we believe to be feasible.
While CN does not present itself as a consume-and-produce tool per se, a close examination of its codebase revealed architecture similar to such tools, so we expect our approach to adapt readily to CN; however, as CN targets only C, there is no strong reason to do so.
We do not know enough about the implementation of Infer-Pulse to assess the feasibility of our approach there.
Perhaps more importantly, we believe our approach will have more impact on the development of new CSE tools than on retrofitting~existing~ones.

\subsubsection*{Scope of the combinator approach (\Cref{sec:zoo} and \Cref{sec:zoo-engine})}
In this paper, our primary focus has been on combinators for programming-language memory models.
We have also introduced two engine combinators, for user-defined predicates and bi-abduction, but these do not cover all CSE engine features: for example, we do not currently have combinators for
\begin{enumerate*}
\item fractional permissions at the level of predicates (as in VeriFast and Viper);
\item predicate constructors, a simple form of higher-order predicates present in VeriFast;
\item magic wands~\cite{lightweight-support-magic-schwerhoff-2015,sound-automation-magic-dardinier-2022} (as in Viper); and
\item iterated permissions~\cite{automatic-verification-iterated-muller-2016} (as in Viper and, to a lesser extent, CN),
\end{enumerate*}
nor have we applied our approach to implicit dynamic frames as used by Viper.
We see no fundamental obstacle to extending our approach to these.

\subsubsection*{Threats to the validity of the empirical evaluation (\Cref{sec:eval})}
Our empirical evaluation compares only against Gillian and replays its existing case studies, so it is only as strong as their coverage (the bi-abduction comparison, in particular, is not like-for-like).
Moreover, as the Rocq mechanisation is not executable, the OCaml combinators are written by hand; our confidence in their correctness rests on the observed behavioural equivalence and on the implementations being small and self-contained.

\subsubsection*{Implementation and proof effort}
Designing a memory model with combinators is akin to designing an RA for Iris using RA combinators~\cite{Jung18}: one identifies the ``flavours of resources'' (exclusive, read-only, fractional, etc.) and how they compose.
\emph{Using} combinators requires no implementation or proof work (beyond occasional glue code; \Cref{sec:eval}), as the implementation and soundness of a composite model follow from those of its components.
\emph{Defining} a new combinator, on the other hand, requires implementation and proof work, which may be arbitrarily difficult, but is a one-time cost; moreover, as each combinator captures a single aspect of the memory model, we found this work smaller and simpler than for monolithic models.
However, picking the correct variant of a given combinator (e.g., of the partial map) currently requires either experimentation~or~experience.

%% file: sections/related-work.tex
\section{Related Work}%
\label{sec:related-work}

There exists an extensive body of work on the modular construction of various artefacts in programming languages (ranging from semantics and computational effects~\cite{Liang95,swierstra08,Xia19} to compilers~\cite{Day13} to build systems~\cite{Mokhov18}). Our work is the first to address modular construction of CSE engines. We compare with our two closest neighbours: previous work on CSE and program logics.

\subsubsection*{CSE} 
In order to be compatible with our à la carte approach, a CSE engine needs to be parametric on the memory model. 
Currently, the only CSE engine that implements this type of parametricity is Gillian, which also supports both OX and UX reasoning. The three most  well-known modern  CSE engines for OX reasoning, VeriFast, Viper, and CN, use instead a hard-coded memory model consisting of a multiset of fractionally or exclusively owned resources and user-defined predicates. 
While the recently introduced CSE engine Raven~\cite{Gupta25} is parametric on ghost state, it still uses the Viper memory model.
The work on these other tools, and theories inspired by these tools, e.g., \citet{featherweight-verifast-jacobs-2015} and \citet{formal-foundations-translational-dardinier-2025}, has therefore not considered modular construction of memory models. 
Apart from Gillian, the only other UX tool is the 
Infer-Pulse platform~\cite{finding-real-bugs-le-2022,unter} that inspired ISL. While Infer-Pulse does analyse many languages, the published papers only discuss simple linear memory, with symbolic execution presented as a proof search within ISL.

\subsubsection*{Program logics} 
As CSE builds on SL, there is overlap to be found in previous work on the modular construction of SLs. 
Work that modularly constructs resource algebras for separation logics, such as that of~\citet{Dockins09} and~\citet{Jung18}, may at first look deceptively similar.
However, the structures that we modularly construct in this paper, specifically concrete and symbolic memory models, contain substantially more data than just the concrete algebraic structures of resource algebras. 
In particular, our structures also contain, among other data, memory actions as well as consume and produce operations which are readily usable inside a CSE engine for automated reasoning, together with accompanying soundness proofs.
Additionally, previous work for program logics only covers separation logic for OX reasoning.
We are not aware of work on modular construction of basic algebraic structures that induce a separation logic for UX reasoning.
More recently, \citet{Vistrup25} have developed a framework for modularly constructing Iris logics by composing different computational effects such as state, non-determinism, and concurrency. Extending our approach to computational effects is an interesting avenue for future work.

%% file: sections/conclusion.tex
\section{Conclusions}
\label{sec:concs}

We have introduced a combinator-based modular approach to building memory models of CSE engines, extending the CSE theory by \citet{compositional-symbolic-execution-loow-2025} in Rocq with combinators and implementing them in the Gillian multi-language platform~\cite{gillian-part-multilanguage-fragososantos-2020,gillian-part-ii-maksimovic-2021,compositional-symbolic-execution-loow-2024}. We have shown  how our approach applies to theoretical memory models, developing simulation equivalences between the combinator-based and monolithic models in Rocq, and to real-world memory models, demonstrating that they exhibit equivalent behaviour and achieve similar performance using a number of previous case studies done using Gillian. Going forward, we hope that combinators will be the canonical approach to building new memory models for CSE tools. 

\section*{Disclosure of Use of AI in Research}

We used Claude (Anthropic) as a coding assistant for two parts of this work: to help write the Rocq definition and soundness proofs of the bi-abduction combinator, and to help refactor the Gillian implementation and its case studies to fit the predicate combinator.

\section*{Data-availability Statement}

The Rocq mechanisation and the Gillian implementation, together with scripts reproducing all case studies, are available in the accompanying artefact~\cite{cse3-artifact}.

\begin{acks}
We thank Dominique Larchey-Wendling and Jean Abou Samra for help with how to approach a Rocq proof with nested recursion for the user-defined-predicate combinator. This work was partially supported by the Amazon Research Award ``Gillian-Rust: Unbounded Verification for Unsafe Rust Code'' (Ayoun), Gardner's faculty gift from Meta (Lööw), and an Imperial College London Department of Computing PhD Scholarship (Park).

\end{acks}

%% file: sections/app-zoo.tex
\section{Additional Definitions for Memory-model Atoms and Combinators}%
\label{app:zoo}

\subsection{Exclusive Ownership}

The $\exclstm(\type)$ MMA captures exclusive ownership of a value of type $\type$. Its concrete variant has as its carrier type the option type $\type?$: that is, either $\Some(v)$, denoting ownership of $v \in \type$, or $\None = \emptycmem$, denoting no ownership. The only defined composition is $\cmem \memcomp \None = \cmem$, with the intuition that only one party can own a value at a time, and composition of two non-empty ownerships is~disallowed.
There are two actions, $\alookup$ and $\amutate$, which, respectively, read and write the owned value. Both actions require ownership of the value to be performed, and yield a $\oxm$ outcome when attempted without ownership. There is a single resource, denoted by $\exclpred(;v)$,
such that $\cmem \resmodels \exclpred(;v) \iff \cmem = \Some(v)$. Note that this resource does not have in-parameters, as there exists only one value that can be retrieved from memory.

The symbolic carrier type is the option type $\SExp?$. The satisfaction relation for symbolic memories is as follows:
\[
\begin{array}{l@{~}c@{~}l}
\subst, \Some(\esem{\sexp}{\subst}) & \memmodels & \Some(\sexp) \\
\subst, \None & \memmodels & \None
\end{array}
\]

The symbolic $\alookup$ and $\amutate$ actions are defined analogously to the concrete ones, and require ownership (i.e., $\smem \neq \None$) to be performed.
Consuming $\exclpred(;\sexp)$ sets the memory to $\None$ and returns $\sexp$ if the symbolic memory is $\Some(\sexp)$, and aborts otherwise.
Producing $\exclpred(;\sexp)$, mirroring concrete composition, yields $\Some(\sexp)$ if the symbolic memory is $\None$, and has no result otherwise.

\subsection{Product Combinator}

The product combinator $\cmm_1 \prodstm \cmm_2$ is a simple but useful combinator that forms the product of two memory models $\cmm_1$ and $\cmm_2$. Important (parts of) definitions of the concrete combinator~include:

\begin{minipage}{0.5\textwidth}\centering%
\[\begin{array}{l@{~}c@{~}l}
\cmemss & \defeq & \cmemss^{\cmm_1} \times \cmemss^{\cmm_2} \\
\cwf((\cmem_1, \cmem_2)) & \defeq & \cwf^{\cmm_1}(\cmem_1) \land \cwf^{\cmm_2}(\cmem_2) \\
(\cmem_1, \cmem_2) \memcomp (\cmem_1', \cmem_2') & \defeq & (\cmem_1 \memcomp^{\cmm_1} \cmem_1', \cmem_2 \memcomp^{\cmm_2} \cmem_2') \\
\emptycmem & \defeq & (\emptycmem^{\cmm_1}, \emptycmem^{\cmm_2}) \\
\actions & \defeq & \inl(\action_1 : \actions^{\cmm_1}) \mid \inr(\action_2 : \actions^{\cmm_2}) \\
\resources & \defeq & \inl(\resource_1: \resources^{\cmm_1}) \mid \inr(\resource_2: \resources^{\cmm_2}) \\
\end{array}\]%
\end{minipage}%
\begin{minipage}{0.5\textwidth}\centering%
\begin{mathpar}\small%
\inferrule[\text{Act-Prod-Left}]{
  \cmem_1.\action_1(\vec{v}) \baction \outcome: (\vec{v}', \cmem_1')
}{
  (\cmem_1, \cmem_2).\inl(\action_1)(\vec{v}) \baction \outcome: (\vec{v}', (\cmem_1', \cmem_2))
}%
\and%
\begin{array}{l}
(\cmem_1, \cmem_2) \models \inl(\resource_1)(\vin; \vout) \\
\qquad\iff \cmem_1 \models \resource_1(\vin; \vout) \land \cmem_2 = \emptycmem^{\cmm_2}%
\end{array}
\end{mathpar}%
\vspace{0.1cm}
\end{minipage}%

In short, concrete memories of $\cmm_1 \prodstm \cmm_2$ are pairs of elements of each memory model, with composition, well-formedness, and the empty memory defined pointwise from the two components. The actions of both components are lifted to the product model using a tagged union; that is, an action of the form $\inl(\action_1)$ (resp. $\inr(\action_2)$) executes action $\action_1$ (resp.~$\action_2$) on the first (resp.~second) component of the memory, leaving the other component unchanged, as illustrated by the \textsc{Act-Prod-Left} rule. Similarly, both sets of resources are lifted and their satisfaction relations are defined pointwise.

The symbolic combinator $\smm_1 \prodstm \smm_2$ is defined analogously to the concrete one, with symbolic memories being pairs of symbolic memories of each component, actions and resources lifted in the same way, the satisfaction relation defined pointwise, and the resource-level consume and produce operations acting on only the relevant component of the memory.

\subsection{Sum Combinator}

The sum combinator, denoted by $\cmm_1 \sumstm \cmm_2$, describes memories that belong to either $\cmm_1$ or $\cmm_2$. In a plot twist of sorts, and for reasons clarified below, the sum memories are defined as the tagged union of component well-formed memories plus a dedicated empty memory $\sumzero$, whereas the well-formed memories exclude the components' empty memories:
\[\begin{array}{l@{~\defeq~}l@{~}l}
\cmemss & \inl(\cmem_1 : \cmemss^{\cmm_1}) \mid \inr(\cmem_2 : \cmemss^{\cmm_2}) \mid \sumzero \\
\cwf & \big((\{ \inl(x) \mid x \in \cwf^{\cmm_1} \} \cup \{ \inr(x) \mid x \in \cwf^{\cmm_2} \}) \setminus \{ \inl(\emptycmem^{\cmm_1}), \inr(\emptycmem^{\cmm_2}) \}\big) \cup \{ \sumzero \} \\
\end{array}\] 

Composition lifts component composition to the tagged union, disallowing composition of memories from different memory models. As for the product combinator, actions/resources are the tagged union of the actions/resources of each component, and their semantics is lifted to operate normally when the memory belongs to the relevant component, and yield an error/abort outcome otherwise. To ensure consistency, all actions and all consume and produce operations that yield an empty memory for either component yield $\sumzero$ instead when lifted to the sum.

To justify the decisions about empty memories made above, consider the sum $\exclstm(\type) \sumstm \exclstm(\type)$, and assume that the components' empty memories are well-formed. 
In the consume-and-produce paradigm of CSE, applying the specification $\tripleok{\inl(\exclpred(;x))}{\textsf{f}()}{\inr(\exclpred(;y))}$ for some function \textsf{f} consists in consuming the pre-condition and producing the post-condition. Consider the memory $\inl(\Some(v))$: consuming $\inl(\exclpred(;v))$ would yield $\inl(\None)$, but from there, producing $\inr(\exclpred(;v'))$ would have no result, as composition is not defined on memories from two different models. This, however, would be unsound, since the specification is valid and applying it should have a result (having none indicates an infeasible path).

%% file: sections/app-eval.tex
\section{Definition of Simulation Relations for Theoretical Evaluation}%
\label{app:app-eval}

We formally define the following simulation relations used for the theoretical evaluation of the OX-only linear memory model, the exact linear memory model with fractional permissions, and the exact block-offset memory model. We first define some common notations used in these simulation relations. 

For concrete simulation $\cmm_1 \sqsubseteq \cmm_2$, assume we have concrete memory models $\cmm_1$ and $\cmm_2$.
\begin{itemize}
    \item Given $\hp_1 \in \cmemss^{\cmm_1}$ and $\hp_2 \in \cmemss^{\cmm_2}$, the relation $\hp_1 \simeq_\cmemss \hp_2$ states that $\hp_1$ and $\hp_2$ are related, e.g., the domains of both memories match exactly, and the corresponding mappings in the memories are also related.
    \item Given $\action_1 \in \actions^{\cmm_1}$ and $\action_2 \in \actions^{\cmm_2}$, the relation $\action_1 \simeq_\actions \action_2$ states that actions $\action_1$ and $\action_2$ are related.
    \item Given $\vinone$ and $\vintwo$, $\vinone \simeq_{\emph{ins}} \vintwo$ states that the inputs for a given concrete action are related.
    \item Given $\voutone$ and $\vouttwo$, $\voutone \simeq_{\emph{outs}} \vouttwo$ states that the outputs for a given concrete action are related.
\end{itemize}

Furthermore, for (compositional) symbolic simulations $\smm_1\ \hat{\sqsubseteq}\ \smm_2$, $\smm_1\ \hat{\sqsubseteq}_\produce\ \smm_2$, and $\smm_1\ \hat{\sqsubseteq}_\mac\ \smm_2$, assume we have symbolic memory models $\smm_1$ and $\smm_2$, extended from $\cmm_1$ and $\cmm_2$, respectively. 

\begin{itemize}
    \item Given $\smem_1 \in \smemss^{\smm_1}$ and $\smem_2 \in \smemss^{\smm_2}$, the relation $\smem_1 \simeq_\smemss \smem_2$ states that $\smem_1$ and $\smem_2$ are related, e.g., the domains of both memories match exactly, and the corresponding mappings in the memories are also related.
    \item Given $\vec{\sexp}_{in1}$ and $\vec{\sexp}_{in2}$, $\vec{\sexp}_{in1} \simeq_{\emph{ins}} \vec{\sexp}_{in2}$ states that the inputs for a given symbolic action, consume resource operation, or produce resource operation are related.
    \item Given $\vec{\sexp}_{out1}$ and $\vec{\sexp}_{out2}$, $\vec{\sexp}_{out1} \simeq_{\emph{outs}} \vec{\sexp}_{out2}$ states that the outputs for a given symbolic action, consume resource operation, or produce resource operation are related.
    \item Given path conditions $\spc_1$ and $\spc_2$, $\spc_1 \simeq_{\spc} \spc_2$ states that the two path conditions are logically equivalent, and both path conditions also use the same set of logical variables.
    \item Given $\resource_1 \in \resources^{\cmm_1}$ and $\resource_2 \in \resources^{\cmm_2}$, the relation $\resource_1 \simeq_\resources \resource_2$ states that $\resource_1$ and $\resource_2$ are related.
\end{itemize}

\subsection{Concrete Simulation Relation}

The formal definition for the concrete simulation relation $M_1 \sqsubseteq M_2$ is given below.

\[
\begin{array}{l}
\hp_1.\alpha_1(\vinone) \baction \outcome: (\voutone, \hp_1')\ \land \\
\hp_1 \simeq_\cmemss \hp_2\ \land \\
\cwf_1(\hp_1) \implies \\
\quad\exists \hp_2', \vouttwo, \alpha_2. \vintwo.\ \\
\quad\quad\action_1 \simeq_\actions \action_2\ \land\\
\quad\quad\vinone \simeq_{\emph{ins}} \vintwo\ \land\\
\quad\quad\hp_1' \simeq_\cmemss \hp_2'\ \land\\
\quad\quad \voutone \simeq_{\emph{outs}} \vouttwo\ \land \\
\quad\quad \hp_2.\alpha_2(\vintwo) \baction \outcome: (\vouttwo, \hp_2')
\end{array}
\]

If both $\cmm_1 \sqsubseteq \cmm_2$ and $\cmm_2 \sqsubseteq \cmm_1$ hold, where the relations used within the concrete simulation relation are inverted, then we say that there is a concrete simulation equivalence relation $\cmm_1 \simeq \cmm_2$.

\subsection{Symbolic Simulation Relation}

The formal definition for the symbolic simulation relation $\smm_1\ \hat{\sqsubseteq}\ \smm_2$ is given below.

\[
\begin{array}{l}
\smem_1.\action(\vec{\sexp}_{in1}, \spc_1) \baction_m \outcome : (\vec{\sexp}_{out1}, \smem_1', \spc_1')\ \land\\
\smem_1 \simeq_{\smemss} \smem_2\ \land\\
\mathsf{SAT}(\spc_1')\ \land\\
\spc_1 \simeq_{\spc} \spc_2\ \implies \\
\quad\exists \hp_2', \spc_2, \vec{\sexp}_{out2}, \alpha_2. \vec{\sexp}_{in2}. \\
\quad\quad\alpha_1 \simeq_\actions \alpha_2\ \land\\
\quad\quad\vec{\sexp}_{\emph{in1}} \simeq_{\vec{\sexp}} \vec{\sexp}_{in2}\ \land\\
\quad\quad\smem_1' \simeq_{\smemss} \smem_2'\ \land\\
\quad\quad\vec{\sexp}_{\emph{out1}} \simeq_{\vec{\sexp}} \vec{\sexp}_{out2}\ \land\\ 
\quad\quad\spc_1' \simeq_{\spc} \spc_2'\ \land \\
\quad\quad\smem_2.\alpha_2(\vec{\sexp}_{in2}, \spc_2) \baction_m \outcome: (\vec{\sexp}_{out2}, \hp_2', \spc_2')
\end{array}
\]

If both $\smm_1\ \hat{\sqsubseteq}\ \smm_2$ and $\smm_2\ \hat{\sqsubseteq}\ \smm_1$ hold, where the relations used within the symbolic simulation relation are inverted, then we say that there is a symbolic simulation equivalence relation $\smm_1\ \hat{\simeq}\ \smm_2$.

\subsection{$\resconsume$ Simulation Relation}

The formal definition for the consume resource simulation relation $\smm_1\ \hat{\sqsubseteq}_{\mac}\ \smm_2$ is given below.

\[
\begin{array}{l}
\smem_1.\resconsumeone(\resource_1, \vec{\sexp}_{in1}, \spc_1) \baction_m \outcome:  (\vec{\sexp}_{out1}, \smem_1', \spc_1')\ \land\\
\smem_1 \simeq_{\smemss} \smem_2\ \land\\
\spc_1 \simeq_{\spc} \spc_2\ \implies \\
\quad\exists \hp_2', \spc_2, \vec{\sexp}_{out2}, \alpha_2. \vec{\sexp}_{in2}. \\
\quad\quad\resource_1 \simeq_\resources \resource_2\ \land\\
\quad\quad\vec{\sexp}_{\emph{in1}} \simeq_{\vec{\sexp}} \vec{\sexp}_{in2}\ \land\\
\quad\quad\smem_1' \simeq_{\smemss} \smem_2'\ \land\\
\quad\quad\vec{\sexp}_{\emph{out1}} \simeq_{\vec{\sexp}} \vec{\sexp}_{out2}\ \land\\ 
\quad\quad\spc_1' \simeq_{\spc} \spc_2'\ \land \\
\quad\quad\resconsumetwo(\resource_2, \vec{\sexp}_{in2}, \spc_2) \baction_m \outcome:  (\vec{\sexp}_{out2}, \smem_2', \spc_2')
\end{array}
\]

If both $\smm_1\ \hat{\sqsubseteq}_{\mac}\ \smm_2$ and $\smm_2\ \hat{\sqsubseteq}_{\mac}\ \smm_1$ hold, where the relations used within the consume resource simulation relation are inverted, then we say that there is a consume resource simulation equivalence relation $\smm_1\ \hat{\simeq}_{\mac}\ \smm_2$.

\subsection{$\resproduce$ Simulation Relation}

The formal definition for the produce resource simulation relation $\smm_1\ \hat{\sqsubseteq}_{\produce}\ \smm_2$ is given below.

\[
\begin{array}{l}
\smem_1.\resproduceone(\resource_1, \vec{\sexp}_{in1}, \vec{\sexp}_{out1}, \spc_1) \baction_m (\smem_1', \spc_1')\ \land\\
\smem_1 \simeq_{\smemss} \smem_2\ \land\\
\spc_1 \simeq_{\spc} \spc_2\ \implies \\
\quad\exists \hp_2, \spc_2, \vec{\sexp}_{out2}, \alpha_2. \vec{\sexp}_{in2}. \\
\quad\quad\resource_1 \simeq_\resources \resource_2\ \land\\
\quad\quad\vec{\sexp}_{\emph{in1}} \simeq_{\vec{\sexp}} \vec{\sexp}_{in2}\ \land\\
\quad\quad\smem_1' \simeq_{\smemss} \smem_2'\ \land\\
\quad\quad\vec{\sexp}_{\emph{out1}} \simeq_{\vec{\sexp}} \vec{\sexp}_{out2}\ \land\\ 
\quad\quad\spc_1' \simeq_{\spc} \spc_2'\ \land \\
\quad\quad\smem_2.\resproducetwo(\resource_2, \vec{\sexp}_{in2}, \vec{\sexp}_{out2}, \spc_2) \baction_m (\smem_2', \spc_2')
\end{array}
\]

If both $\smm_1\ \hat{\sqsubseteq}_{\produce}\ \smm_2$ and $\smm_2\ \hat{\sqsubseteq}_{\produce}\ \smm_1$ hold, where the relations used within the produce resource simulation relation are inverted, then we say that there is a produce resource simulation equivalence relation $\smm_1\ \hat{\simeq}_{\produce}\ \smm_2$.

\subsection{Simulation Equivalence for the OX Linear Memory Model}

Let $M_{\emph{LinOXCom}}$ be the combinator-built linear memory model constructed using the following combinators: $\pmapstm(\nats, \freeableoxstm(\exclstm(\val)))$, and let $M_{\emph{LinOX}}$ be the monolithic OX-only linear memory model.
\begin{itemize}
    \item Given $\hp_1 \in \cmemss^{\cmm_{\emph{LinOXCom}}}$ and $\hp_2 \in \cmemss^{\cmm_{\emph{LinOX}}}$, the relation $\hp_1 \simeq_\cmemss \hp_2$ is formally defined as: 
    \[
    \begin{array}{l}
        \dom(\mu_1) = \dom(\mu_2) \land (\forall a.\ a \in \dom(\mu_1) \implies \mu_1(a) = \Some (\mu_2(a)))
    \end{array}
    \]
    \item Given $\action_1 \in \actions^{\cmm_{\emph{LinOXCom}}}$ and $\action_2 \in \actions^{\cmm_{\emph{LinOX}}}$, the relation $\action_1 \simeq_\actions \action_2$ states that actions $\action_1$ and $\action_2$ are related as follows:
    \begin{itemize}
        \item when $\action_1 = \anew$ (from the $\pmapstm$ combinator) and $\action_2 = \anew$,
        \item when $\action_1 = \asub(\afree)$ and $\action_2 = \afree$,
        \item when $\action_1 = \asub(\asub(\amutate))$ and $\action_2 = \amutate$, and
        \item when $\action_1 = \asub(\asub(\alookup))$ and $\action_2 = \alookup$.
    \end{itemize}
    \item The relation $\vinone \simeq_{\emph{ins}} \vintwo$ is defined to be a strict equality (and also for symbolic simulation, which takes in symbolic expressions instead).
    \item The relation $\voutone \simeq_{\emph{outs}} \vouttwo$ is defined such that every error in the former model has a corresponding unique match in the latter model (and also for symbolic simulation, which takes in symbolic expressions instead).
    \item Given $\smem_1 \in \smemss^{\smm_{\emph{LinOXCom}}}$ and $\smem_2 \in \smemss^{\smm_{\emph{LinOX}}}$, the relation $\smem_1 \simeq_\smemss \smem_2$ is defined as a simple direct lifting of its concrete variant (i.e., $\simeq_\cmemss$).
    \item Given $\resource_1 \in \resources^{\cmm_{\emph{LinOXCom}}}$ and $\resource_2 \in \resources^{\cmm_{\emph{LinOX}}}$, the relation $\resource_1 \simeq_\resources \resource_2$ states that $\resource_1$ and $\resource_2$ are related when $\resource_1 =\ \exclpred$ (from the $\exclstm$ MMA) and $\resource_2 =\ \exclpred$.
\end{itemize}
We have shown that the following simulation relations hold:
\begin{itemize}
    \item $\cmm_{\emph{LinOXCom}} \sqsubseteq \cmm_{\emph{LinOX}}$
    \item $\smm_{\emph{LinOXCom}}\ \hat{\sqsubseteq}\ \smm_{\emph{LinOX}}$
    \item $\smm_{\emph{LinOXCom}}\ \hat{\sqsubseteq}_\produce\ \smm_{\emph{LinOX}}$
    \item $\smm_{\emph{LinOXCom}}\ \hat{\sqsubseteq}_\mac\ \smm_{\emph{LinOX}}$
\end{itemize}
We have also shown that the inverse relations hold, where all the internal relations have also been inverted:
\begin{itemize}
    \item $\cmm_{\emph{LinOX}} \sqsubseteq \cmm_{\emph{LinOXCom}}$
    \item $\smm_{\emph{LinOX}}\ \hat{\sqsubseteq}\ \smm_{\emph{LinOXCom}}$
    \item $\smm_{\emph{LinOX}}\ \hat{\sqsubseteq}_\produce\ \smm_{\emph{LinOXCom}}$
    \item $\smm_{\emph{LinOX}}\ \hat{\sqsubseteq}_\mac\ \smm_{\emph{LinOXCom}}$
\end{itemize}
As we have shown that the simulation relations and their inverses hold, the following behavioural equivalence relations also hold as corollaries:
\begin{itemize}
    \item $\cmm_{\emph{LinOXCom}} \simeq \cmm_{\emph{LinOX}}$
    \item $\smm_{\emph{LinOXCom}}\ \hat{\simeq} \smm_{\emph{LinOX}}$
    \item $\smm_{\emph{LinOXCom}}\ \hat{\simeq}_\produce\ \smm_{\emph{LinOX}}$
    \item $\smm_{\emph{LinOXCom}}\ \hat{\simeq}_\mac\ \smm_{\emph{LinOX}}$
\end{itemize}

From~\Cref{sec:eval-theory}, these two models are behaviourally equivalent up to order of error checking and branching. This is for two reasons:
\begin{itemize}
    \item the combinator-built model fixes the order of error checking in the semantics;
    and
    \item for symbolic execution, the combinator-built model uses a fixed branching strategy.
\end{itemize} 
We made some changes in $\cmm_{\emph{LinOX}}$ to the order of error checking in the input arguments of the actions, and, for the symbolic model, we made changes to the branching strategies such that the strategy exactly matches that of the combinator-built model.

To give an example of the former point, consider the first few cases of the concrete $\amutate$ semantics in the original OX-only linear memory model (before we made the changes):
\begin{mathpar}
\small
\inferrule
{|\vec{v}| \neq 2}
{\hp.\amutate(\vec{v}) \baction^{\emph{LinOX}} \oxerr: ([``\mathsf{length}"], \hp)}
\and
\inferrule
{v_a \notin \nats}
{\hp.\amutate([v_a, v]) \baction^{\emph{LinOX}} \oxerr: ([``\mathsf{type}"], \hp)}
\end{mathpar}
Note that the $\amutate$ rule checks whether there are exactly two arguments before checking the type of the first argument.
However, the combinator-built model performs these checks combinator by combinator. The $\pmapstm$ combinator first checks the existence of the first argument and then type checks the first argument. Afterwards, the $\exclstm$ MMA checks the existence of the second argument and checks that there are no further arguments. Due to this particular order of error checking, we made the following changes to the above cases of the $\amutate$ semantics in the original OX-only linear memory model:

\begin{mathpar}
\small
\inferrule
{\ }
{\hp.\amutate([]) \baction^{\emph{LinOX}} \oxerr: ([``\mathsf{length}"], \hp)}
\and
\inferrule
{v_n \notin \nats}
{\hp.\amutate(v_n :: \vec{v}) \baction^{\emph{LinOX}} \oxerr: ([``\mathsf{type}"], \hp)}
\and
\inferrule
{n \in \nats \land |\vec{v}| \neq 1}
{\hp.\amutate(n :: \vec{v}) \baction^{\emph{LinOX}} \oxerr: ([``\mathsf{length}"], \hp)}
\end{mathpar}

Making these changes, we were able to show behavioural equivalence for the $\amutate$ rule. The same idea applies to all other rules.

We note that only the case of checking the well-formedness of the input arguments was affected, even for the other simulation-related models such as the exact block-offset memory model. In short, this shows the successful cases matched exactly, and the remaining cases also matched under the assumption that a well-formed input was provided. 
 
Now we give an example regarding the latter point. Consider the following case in the symbolic $\alookup$ semantics in the original OX-only linear memory model (before we made the changes):

\begin{mathpar}
\small
\infer
 {\smem.\alookup(\vec{\sexp}, \spc) \baction_m^{\emph{LinOX}} \oerr : ([``\mathsf{length}"], \smem, \spc)}
 {|\vec{\sexp}| \neq 1}
\end{mathpar}

Here, in addition to directly checking whether there exists exactly one argument as with the $\amutate$ case above, this case checks whether there are more input arguments than necessary before checking whether a mapping from $\sexp_a$ exists in memory $\smem$ (and afterwards, the action would check whether the type matches and whether the mapping exists). However, for the combinator-built linear memory model, the lookup rule first checks whether the type matches but, more importantly, branches on whether $\sexp_a \notin \dom(\smem)$ holds or not. These differences in branching strategies affect proving the simulation relation, so we made some slight modifications to this case in the monolithic memory model, where there are two branches based on the existence of the mapping:
\begin{mathpar}
\small
\inferrule
 {\vec{\sexp} \neq []
  \\\\
  \spc' \defeq \spc \land (\sexp_a \in \nats) \land 
  (\sexp_a \notin \dom(\smem))}
 {\smem.\alookup(\sexp_a :: \vec{\sexp}, \spc) \baction_m^{\emph{LinOX}} \oerr : ([``\mathsf{length}"], \smem, \spc')}
\and
\inferrule
 {\vec{\sexp} \neq []
  \and
  \smem(\sexp_a') = \sexp_v
  \\\\
  \spc' \defeq \spc \land (\sexp_a \in \nats) \land
  (\sexp_a = \sexp_a')}
 {\smem.\alookup(\sexp_a :: \vec{\sexp}, \spc) \baction_m^{\emph{LinOX}} \oerr : ([``\mathsf{length}"], \smem, \spc')}
\end{mathpar}

The modifications are minor in that only the order of checking the well-formedness of the input of actions was affected. In other words, we have shown that if we assume we have a well-formed input for actions, then the rest of the semantics of the two models matches exactly. Consequently, this means the successful cases of the original model were left untouched, and the semantics matched exactly.